\documentclass[12pt,letterpaper]{article}  
                
\usepackage[includeheadfoot, 
            marginratio={1:1,2:3}, 
            width=412pt, 
            height=688pt]{geometry}

\usepackage{amsmath}
\usepackage{amsfonts}
\usepackage{amssymb}

\usepackage{graphicx}
\usepackage[margin=10pt,font=small]{caption}
\usepackage{mathtools}

\newcommand{\ov}{\overline}
\newcommand{\eq}[1]{\begin{equation}
                     \begin{split} #1 \end{split}
                     \end{equation}}
\newcommand{\tri}{\hspace{-3.5pt}\vartriangle\hspace{-3.5pt}}

\numberwithin{equation}{section}

\begin{document}

\vspace*{-1.5cm}
\begin{flushright}
  {\small
  MPP-2011-59\\
  LMU-ASC 17/11\\
  ITP-UU-11/20\\
  SPIN-11/15\\ 
  }
\end{flushright}

\vspace{1.25cm}
\begin{center}
  {\LARGE
Non-geometric Fluxes, Asymmetric Strings and\\[0.2cm]
Nonassociative Geometry  \\[0.2cm]
  }
\end{center}

\vspace{0.5cm}
\begin{center}
  Ralph Blumenhagen$^{1}$, Andreas Deser$^{1}$, Dieter L\"ust$^{1,2}$, \\[0.1cm]
  Erik Plauschinn$^{3}$ and Felix Rennecke$^{1}$ 
\end{center}

\vspace{0.1cm}
\begin{center} 
\emph{$^{1}$ Max-Planck-Institut f\"ur Physik (Werner-Heisenberg-Institut), \\ 
   F\"ohringer Ring 6,  80805 M\"unchen, Germany } \\
\vspace{0.25cm} 
\emph{$^{2}$ Arnold Sommerfeld Center for Theoretical Physics,\\ 
               LMU, Theresienstr.~37, 80333 M\"unchen, Germany}\\
\vspace{0.25cm}
\emph{$^{3}$ Institute for Theoretical Physics and Spinoza Institute, \\
Utrecht University, 3508 TD  Utrecht, The Netherlands}  \\

\end{center} 

\vspace{1cm}


\begin{abstract}
We study closed bosonic strings propagating both in a flat background with constant  $H$-flux and in its T-dual configurations. We define a conformal field theory capturing linear effects in the flux and compute scattering amplitudes of tachyons, where the Rogers dilogarithm plays a prominent role. For the scattering of four tachyons, a fluxed version of the Virasoro-Shapiro amplitude is derived and its pole structure is analyzed.  
In the case of an $R$-flux background obtained after three T-dualities, we find indications for a nonassociative target-space structure which can be described in terms of a deformed tri-product. Remarkably, this product is compatible with
crossing symmetry of conformal correlation functions.
We finally argue that the $R$-flux background flows to an asymmetric CFT.
\end{abstract}


\clearpage

\tableofcontents


\section{Introduction}
\label{sec:intro}

The basic space-time principles underlying string theory
are still elusive to a large extend.
Indeed, one of the important differences between
a string and a point particle is how they probe the space-time 
they are moving in. For instance, due to the decoupling of 
the left- and right-moving sectors, strings can not only propagate
in left-right symmetric and thus geometric backgrounds, 
but also in backgrounds defined via a left-right asymmetric conformal field
theory. Examples are   asymmetric orbifolds,
whose target-space interpretation is not clear a priori. Note that for point particles, this situation does 
not occur.
Related to this feature is the observation that strings 
cannot distinguish between spaces which are related by T-duality,
which from the world-sheet point of view is a left-right asymmetric 
operation.

In the past, applying T-duality to known configurations has 
led to  new insight into string theory, where a prominent
example is the discovery of D-branes. Furthermore, in 
\cite{Shelton:2005cf,Wecht:2007wu} T-duality was applied to a
simple closed string background, namely to a flat space with non-vanishing 
three-form flux $H=dB$, which resulted in a background with geometric 
flux. This so-called  twisted torus is still a conventional string background,
but a second T-duality leads to a  non-geometric flux background.
These are spaces in which  the
transition functions between two charts of a manifold are allowed
to be T-duality transformations, hence they are also called T-folds \cite{Dabholkar:2002sy,Hellerman:2002ax,Hull:2004in}. 
After formally applying a third T-duality, not along an isometry direction any more,
one obtains an $R$-flux background which does not 
admit a clear target-space interpretation. It was proposed 
not to  correspond to an ordinary geometry even locally,
but instead to give rise to a nonassociative geometry \cite{Bouwknegt:2004ap,Bouwknegt:2004tr}
(see also \cite{Frohlich:1993es,Lizzi:1997em}).

From another point of view, the general expectation is that in quantum gravity 
a noncommutative structure of space-time emerges. In string theory,
noncommutativity can be  seen for open strings
whose end-points probe a D-brane endowed with a non-trivial
two-form background. In this case, it is a well-established
result that the coordinates on the D-brane do not commute, that is
$[x^a,x^b]=\theta^{ab}$.
Moreover, it has been shown explicitly in \cite{Seiberg:1999vs} that 
one can reformulate the theory on the D-brane
in terms of a noncommutative gauge theory, where the ordinary
product of two fields is deformed to the Moyal-Weyl
product. This structure can
be tested directly by the evaluation of on-shell open string
scattering amplitudes.
The question now is whether a similar noncommutative space-time
structure also appears for closed strings, i.e. for
those objects which correspond to gravity.

In \cite{Blumenhagen:2010hj} it was argued that, since the world-sheet of a closed string
is a sphere, in this case not two but  three coordinates should be involved. 
For the well-understood 
 background of the $SU(2)_k$ WZW model,
the world-sheet equal-time, equal-space
Jacobi-identity for three space-time coordinates was evaluated 
and  found to be non-vanishing. This gave rise
to the conjecture that the space-time coordinates
satisfy a non-trivial three-bracket relation   
$[x^a,x^b,x^c]=\theta^{abc}$, where $\theta^{abc}$
is related to the three-form flux. 
Thus, the space-time
coordinates are not just noncommutative but also nonassociative, and
we call such geometries noncommutative/nonassociative (NCA). 
Let us note that an analogous  structure also appears for an  open string
ending on a D-brane with constant $H$-flux, which was demonstrated
in \cite{Cornalba:2001sm,Herbst:2001ai,Herbst:2002fk}.

In a complementary approach \cite{Lust:2010iy}, the commutation relations between
coordinates of three-dimensional backgrounds with geometric as well as with T-dual non-geometric flux were studied. It was found that for non-geometric flux the commutator between two coordinates $x^a$ and $x^b$ is non-vanishing, and in the case of geometric flux the commutator between  $x^a$ and one dual coordinate $\tilde x^b$ is non-zero (see \cite{Lust:2010iy} for more details).
Studying then the canonical commutation relations between (dual) coordinates and (dual) momenta, one is again led to a nonassociative algebra.

\pagebreak
In this paper, we study some non-geometric aspects of
string theory, in particular asymmetric string solutions, non-geometric
fluxes and nonassociative geometry.
Similar to \cite{Shelton:2005cf,Wecht:2007wu}, our starting point is that of
a closed bosonic string moving through a flat background with  
constant $H$-flux. 
From the string equations of motion one infers
that this configuration is not conformal beyond linear order
in the flux. However, since the three-bracket relation mentioned above is linear
in the flux parameter $\theta^{abc}$, we expect to see NCA effects already at this order. 
We then proceed in the following way.
In section \ref{sec_open},  we briefly review
some results of \cite{Seiberg:1999vs} concerning open string correlators and the derivation
of the noncommutative Moyal-Weyl product, which we want to generalize to the closed string sector.

In section \ref{sec_flux_background}
we construct the conformal field theory  corresponding
to the $H$-flux background at linear order. This involves two effects:
first, the coordinates and related currents are redefined
at linear order in $H$, and second that correlation functions and
operator product expansions (OPEs) receive corrections which 
we evaluate using  conformal perturbation theory.
We compute 
the basic three-point function $\langle \mathcal X^a \mathcal X^b \mathcal X^c\rangle$
of three coordinates $\mathcal X^a$,
which turns out to be proportional to the three-form flux. 
We then introduce vertex operators for the deformed theory and analyze whether
they still correspond to physical states. Here, logarithmic terms appear 
in OPEs and correlation functions.
Furthermore, we analyze T-dual configurations and illustrate that one can
draw direct conclusions on the underlying target-space structure  only for the two cases of pure momentum-mode scattering in the $H$-flux and
in the $R$-flux background.

Section \ref{sec_threep} is devoted to the computation of tachyon scattering
amplitudes.
We find that tachyon amplitudes
are crossing symmetric after  momentum conservation has been employed.
Furthermore, in case of $R$-flux, we detect non-trivial phase
factors which can be encoded in  a  new
nonassociative tri-product.
Then, we analyze the four-tachyon scattering
amplitude in more detail and derive 
a conformally invariant, crossing symmetric fluxed version of the 
Virasoro-Shapiro amplitude.
Like the Veneziano or the usual Virasoro-Shapiro
amplitude,   its pole structure reveals  information on
the spectrum and couplings of the full theory. 
We find new tachyons, indicating instabilities
due to higher order corrections in the flux.

In section \ref{sec_ab_nassgeom} we speculate
how these instabilities lead to new backgrounds after tachyon condensation. 
In particular, we make a proposal
for the case of  $R$-flux which involves a left-right
asymmetric WZW-type background. We also generalize the 
noncommutative Moyal-Weyl product
to a nonassociative tri-product.
Finally, in three appendices we provide additional details on 
the  Rogers dilogarithm,
the (non-)geometry of the T-dual backgrounds and on the 
computation of scattering amplitudes.

\bigskip
Note added in proof: upon finishing this manuscript, a paper with partial overlap compared to our work has appeared \cite{Aldi:2011cf}, which discusses correlation functions and T-duality for backgrounds with geometric flux.


\section{Open and closed strings in flux backgrounds}
\label{sec_open}

In this paper, we  study  closed strings 
 in backgrounds with constant $H$-flux as well as in 
its T-dual configurations. Due to the 
back-reaction of the flux on the metric, this task is
rather difficult. 
However, for open strings ending on a D-brane
with constant two-form background one can analyze
the system via methods of two-dimensional conformal field
theory. In this section, we therefore first recall some basic facts from the  open
string story, before applying similar techniques to the closed string sector in the sequel.


\subsection{The Moyal-Weyl product for  open strings}
\label{sec_review_SW}

We start by reviewing some aspects of the work of
Seiberg and Witten \cite{Seiberg:1999vs} about noncommutativity for open
strings \cite{Connes:1997cr,Schomerus:1999ug}.
In particular, we consider open strings ending on a D-brane 
carrying a non-trivial two-form flux ${\mathcal F}=B+F$. In this case, the two-point function of two open string coordinates $X^a(z)$ inserted on the boundary of a disc  takes the form \cite{Fradkin:1985qd,Callan:1986bc,Abouelsaood:1986gd,Seiberg:1999vs}
\begin{equation}
  \label{twopointa}
   \bigl\langle X^a (\tau_1)\, X^b(\tau_2) \bigr\rangle =  
  - \alpha' G^{ab} \log (\tau_1-\tau_2)^2 + 
   i \, \theta^{ab}\,  \epsilon(\tau_1-\tau_2)\;,
\end{equation}
where $\tau$ stands for the real part of the complex world-sheet coordinate $z$. The matrix $G^{ab}$ is symmetric and can be interpreted as the (inverse of the) effective metric seen by the open string. The matrix $\theta^{ab}$ is  proportional to the two-form flux $\mathcal F$ and thus  is anti-symmetric, and it can be interpreted as a noncommutativity parameter.
The function $\epsilon(\tau)$ is defined as
\eq{
  \epsilon(\tau) = \left\{ \begin{array}{c@{\hspace{20pt}}c}
  +1 & \tau \geq 0 \;, \\
  -1 & \tau <0 \;,
  \end{array} \right.
}  
and it is  the appearance of the jump given by $\epsilon(\tau_1-\tau_2)$ in \eqref{twopointa} which leads to  noncommutativity of the open string coordinates on the D-brane.

Next, we recall the form of open string vertex operators which are inserted at the boundary of a disc diagram.  Employing the short-hand notation $p\cdot X = p_a X^a$ and denoting normal ordered products by $:\!\ldots\!:$, a tachyon vertex operator can be written as
\begin{equation}
  \label{vertexa}
  T\equiv V_{p}(\tau)= \::\! \exp\bigl(\,i\hspace{0.5pt}p \cdot X(\tau)\, \bigr)\!: \;.
\end{equation}
A correlation function of $N$ such vertex operators is found to be
\eq{
  \label{correla}
  \bigl\langle \,T_1\, \ldots T_N \bigr\rangle 
  =\exp\biggl(  i  \sum_{1\le n<m\le N} 
    p_{n,a} \,\theta^{ab}\,   p_{m,b}\,  \epsilon(\tau_n-\tau_m)\biggr)\times
    \bigl\langle \,T_1\, \ldots T_N \bigr\rangle_{\theta=0} \;,
}
which contains an extra phase due to the noncommutative nature of the theory.
Note that because of momentum conservation $\sum_{n=1}^N p_n=0$,
this correlator is invariant under cyclic permutations of the
$N$ vertex operators. Therefore, since a conformal $SL(2,\mathbb R)$ transformation
can only induce  cyclic permutations of  points along
the real axis, the correlation function \eqref{correla}
is $SL(2,\mathbb R)$ invariant. 
One can  then define  an $N$-product $\star_N$ in the following way
\eq{
  \label{Nbracketcon}
  & f_1(x)\, \star_N\,  f_2(x)\, \star_N \ldots \star_N\,  f_N(x) := \\
  &\hspace{40pt}\exp\left( i  \sum_{1\leq n< m\leq N}
     \theta^{ab}\,
      \partial^{x_n}_{a}\,\partial^{x_m}_{b}  \right)\, f_1(x_1)\, f_2(x_2)\ldots
   f_N(x_N)\Bigr|_{x_1=\ldots = x_N=x} \;,
}
which correctly reproduces the phase appearing in \eqref{correla}. 
Note that these $N$-products are  related to 
the subsequent application of the usual star-product $\star=\star_2$ 
\begin{equation}
   f_1\star_N f_2\star_N \ldots \star_N  f_N=
   f_1\star f_2\star \ldots \star  f_N\; ,
\end{equation}
and therefore, by evaluating correlation functions of vertex operators
in open string theory, it is possible to 
derive the Moyal-Weyl product and some of its features.
In the following sections, we apply this approach to correlation functions in the 
closed string sector.

Let us also  recall how to compute 
correlators of massless states such as gluons.
Their (integrated) on-shell vertex operator is given by 
\begin{equation}
  \label{vertexb}
   \mathcal A 
   \equiv
   \int d\tau \,V_{\xi}(\tau)
  =  \int d\tau\,  \xi\cdot \partial_\tau X\, \exp\big( \,i\hspace{0.5pt}p \cdot  X(\tau)\bigr)  \;,
\end{equation}
subject to the restrictions $p^2=0$ and $\xi\cdot p=0$.
The  correlation function of three gluons for a given 
order 
of insertions of  operators along the real axis 
can be computed as
\begin{equation}
\label{gluon3}
\begin{split}
  &\bigl\langle \mathcal A_1\,\mathcal A_2\,\mathcal A_3\bigr\rangle= 
  \Bigl[(\xi_1\cdot \xi_2)(p_2\cdot \xi_3) +   (\xi_1\cdot \xi_3)(p_1\cdot \xi_2) +  
   (\xi_2\cdot \xi_3)(p_3\cdot \xi_1) \\
&  \hspace{40pt} +2\alpha'\, (p_1\cdot \xi_2)(p_2\cdot \xi_3)(p_3\cdot \xi_1)\Bigl]
  \, \exp\Bigl(\, i\, p_{1,a} \,
    \theta^{ab}\, p_{2,b}\, \Bigr)\, \delta(p_1+p_2+p_3)   \;,
\end{split}
\end{equation}
where momentum conservation has been used in the phase factor.
Let us emphasize that from  \eqref{gluon3} we see
that  the noncommutative product 
directly effects gluon scattering
amplitudes, and therefore will modify the corresponding low-energy effective action.


\subsection{Closed strings in flux backgrounds}

In the previous subsection, we have illustrated how to derive the noncommutative Moyal-Weyl product from correlation functions in open string theory, and we have  pointed out  that this product effects scattering amplitudes  and therefore the low energy effective description of the gauge theory.
The question we would like to ask in this paper is whether  this  intriguing
structure also appears in closed string theory.

One of the main differences between the open and closed string in this respect 
is that for the latter 
a constant two-form potential $B$ can be gauged away and the relevant
flux background is expected to be given by the three-form $H=dB$.\footnote{Open
  strings in such an $H$-flux background have been studied for instance in \cite{Alekseev:1999bs,Cornalba:2001sm,Herbst:2001ai,Herbst:2002fk}.}
However, the string equations of motion imply that a non-vanishing
$H$-flux back-reacts on the metric so that a flat background endowed 
with flux does not correspond
to a two-dimensional CFT on the string
world-sheet. On the other hand, as we will review below, the back-reaction
appears only at second order in the flux and we can expect a bona-fide
conformal field theory up to linear order in $H$. 
We denote this theory by CFT$_H$ which we study  in detail in
section \ref{sec_flux_background}.

Let us also comment on configurations T-dual to the above background and 
consider a three-dimensional compact space with constant $H$-flux and vanishing curvature.  
Following the same spirit as in \cite{Shelton:2005cf},  we can
T-dualize this background along an arbitrary number
of directions which is summarized in table
\ref{tableTdual}. There, we have shown  how momentum and winding modes are
exchanged (the mapping for the $Q$- and $R$-flux has only been obtained by
generalizing the $\omega$-flux result), and  more details on these T-dual
backgrounds are collected  in appendix \ref{Tdualbackgrounds}. 
\begin{table}[t]
\centering
\renewcommand{\arraystretch}{1.2}
\begin{tabular}{|c||c|c|c|c|}
\hline
T-dualized directions&  $-$ & $(x^3)$ & $(x^2,x^3)$ & $(x^1,x^2,x^3)$ \\
\hline
T-dual flux  & $H$-flux & $\omega$-flux & $Q$-flux &  $R$-flux \\
\hline
momentum/winding &  $-$ & $(p_3\leftrightarrow w_3)$ & $\Bigl(\begin{matrix}
  p_2\leftrightarrow w_2 \\[-0,2cm] p_3\leftrightarrow w_3 \end{matrix}\Bigr)$  & 
  $\biggl(\begin{matrix}  p_1\leftrightarrow w_1\\[-0,25cm]
  p_2\leftrightarrow w_2 \\[-0,25cm] p_3\leftrightarrow w_3 \end{matrix}\biggr)$\\
\hline
\end{tabular}
\caption{\small T-dualities and their action on the flux and on momentum/winding states.\label{tableTdual}} 
\vspace*{-4pt}
\end{table}
From  table \ref{tableTdual} we see that a first T-duality, say along the $x^3$-direction, maps the $H$-flux to a geometric flux $\omega$ which in the compact case defines a twisted torus. 
After a second T-duality one obtains a so-called T-fold \cite{Dabholkar:2002sy,Hellerman:2002ax,Hull:2004in} defined via a non-geometric $Q$-flux.
The resulting T-fold actually does not any longer posses an isometry
direction so that it is not clear whether a formal T-duality along a
third direction is allowed. It was argued that the resulting 
$R$-flux background is not even locally an ordinary space, but rather
gives rise to a nonassociative geometry \cite{Bouwknegt:2004ap,Bouwknegt:2004tr}.


\section{Conformal field theory with $H$-flux}
\label{sec_flux_background}

In this section, we approach the question whether a noncommutative structure analogous to Seiberg/Witten can be obtained for the closed string.  
To do so, we first  recall some previous results on that matter and  specify the framework for this paper. We then determine the correlator of three closed string coordinates as well as  corrections to the two-point function due to a background $H$-flux. Finally, we introduce  vertex operators for the perturbed theory and discuss  features thereof.


\subsection{Prerequisites}
\label{sec_prereq}

The motivation for our study is the recent paper  \cite{Blumenhagen:2010hj}
where the  question  mentioned above
was discussed 
for the $SU(2)_k$ WZW model.  This model
describes a  string moving on $S^3$
with flux through the sphere, where the radius of $S^3$ is related to the flux such that
 the string equations of motion (corresponding to conformal symmetry
on the world-sheet) are satisfied to all orders in sigma-model perturbation theory. 
For this background, the equal-time cyclic double-commutator of three local coordinates was  found to be  \cite{Blumenhagen:2010hj}
\begin{equation}
  \label{result}
   \lim_{z_i\to z} \Bigl[X^a(z_1,\ov z_1),\bigl[X^b(z_2,\ov
       z_2),X^c(z_3,\ov z_3) \bigr]\Bigr] +{\rm cycl.} 
   =\begin{cases}  \; \epsilon\hspace{0.5pt} \theta^{abc} \quad& z_i\to z \;, \\
                    \;0   & {\rm else}\;, \end{cases}
\end{equation}
where $\theta^{abc}\sim H^{abc}$ encodes the three-form flux
and where the parameter $\epsilon$ turns out to be 
$\epsilon=0$ for the $H$-flux background 
and $\epsilon=1$ for the background one
obtains after an odd number of T-dualities. 
Note that the zero-mode contribution in the above computation was fixed by the
requirement that the Jacobi-identity vanishes if
some of the $z_i$ are not equal (see \cite{Blumenhagen:2010hj} for more details).
In particular,  a constant term which does not depend on $z_i$ is absent in \eqref{result}.
One can then define and compute
\begin{equation}
  \label{result2}
   \bigl[X^a,X^b,X^c \bigr] 
   := \lim_{z_i\to z} \Bigl[X^a(z_1,\ov z_1),\bigl[X^b(z_2,\ov
       z_2),X^c(z_3,\ov z_3) \bigr]\Bigr] +{\rm cycl.} 
   =  \epsilon\, \theta^{abc}\; .
\end{equation}
Thus, in the case of $R$-flux with $\epsilon=1$, the space-time coordinates $X^a$  
satisfy a non-vanishing three-bracket,\footnote{A noncommutative quantum field theory based on a similar three-product has been proposed in \cite{Savvidy:2002gs}.}
and therefore may give rise to a nonassociative geometry.


\subsubsection*{$H$-flux background}

In this paper, we do not start from an exactly solvable
WZW model and then take a local limit, but rather
from a flat background with constant $H$-flux.
This approach is analogous to that for the open string
in a constant $B$-field background, with the main difference  that
here we cannot decouple gravity and so the
back-reaction of the flux has to be considered.

More concretely, our framework is that of a flat space with constant $H$-flux
and dilaton 
which is to be considered as part of a full bosonic string theory construction. 
The metric and the flux are specified by 
\eq{ 
  \label{setup_01}
   ds^2=\sum_{a=1}^N \bigl(dX^a\bigr)^2, \hspace{50pt} 
   H=  \frac{2}{{\alpha'}^2}\,  \theta_{abc}\,  dX^a\wedge dX^b\wedge dX^c\; ,
} 
where in the following we focus mostly  on  $N=3$, but our discussion can be
readily generalized. \pagebreak[2]
Note that already at lowest order in $\alpha'$, this background is not a solution
to the string equations of motion. 
In particular, the beta-functional for the graviton
\eq{
  \label{targeteom}
  \beta_{ab}^G  =  \alpha' R_{ab}-\frac{\alpha' }{4}\: H_{a}{}^{cd}\, H_{bcd}  
  +2\alpha' \nabla_a\nabla_b\Phi
    +O({\alpha'}^2) 
}
does not vanish for \eqref{setup_01} in the case of a constant dilaton $\Phi$. 
Only at {\em linear order} in the $H$-flux 
the above background  provides a solution, and  by dimensional
analysis it is  clear that at higher orders in $\alpha'$ there
can be no further obstructions at linear order in $H$.
We can thus conclude that the flat space background with 
constant $H$ and $\Phi$ corresponds to a 
bona fide conformal field theory at linear order in the flux.
Furthermore,  since the
three-bracket \eqref{result} is linear in  $\theta^{abc}\sim H^{abc}$,
up to first order in the $H$-flux we expect to find 
a reliable world-sheet CFT framework 
capturing potential nonassociative effects.

Now, a closed string moving in the background given by \eqref{setup_01} can be described by a  sigma-model. With $\Sigma$ denoting the world-sheet of the closed string, its action reads
\eq{
  \label{action_730178}
  \mathcal S=\frac{1}{2\pi\alpha'}\int_\Sigma d^2 z\, \bigl(\,
  g_{ab} + B_{ab} \bigr)\, \partial X^a \,\ov\partial X^b
  \;,
}
where the metric for our particular situation is given by $g_{ab} = \delta_{ab}$. For the $B$-field we can choose a gauge in which $B_{ab} = \frac{1}{3}  H_{abc}\, X^c$; and the equations of motion  are of the form
\eq{
  \label{eom_619}
  \partial\hspace{0.5pt} \ov\partial\hspace{0.5pt} X^a = 
  {\textstyle \frac{1}{2}} H^a{}_{bc} \,\partial X^b\, \ov\partial X^c \;.
}
At zero order in $H$, the solution to \eqref{eom_619} is that of the well-known free theory  for which  we employ the notation
\eq{
   \mathsf X^a_0(z,\ov z) =  \mathsf X^a_L(z) + \mathsf{X}{}^a_R(\ov z) \;,
}  
where we made a distinction between the solution $\mathsf X^a$ and the field $X^a$.
At linear order in the flux, a solution to  \eqref{eom_619} is  given by 
\eq{
  \label{def_x_04}
  \mathsf{X}_1^a(z,\ov z) =  \mathsf X_0^a(z,\ov z) + 
  {\textstyle \frac{1}{2}}\hspace{0.5pt}H^a{}_{bc} \,  \mathsf X^b_L(z) \, 
  \mathsf {X}{}^c_R(\ov z)  \;.
}


\subsubsection*{Perturbation theory}

The natural approach  to compute the  correlation functions in the above setting is conformal perturbation theory
 \cite{AlvarezGaume:1981hn,Fradkin:1985qd,Braaten:1985is,Zamolodchikov:1986gt,
Abouelsaood:1986gd,Callan:1986bc,Ludwig:1987gs,Cornalba:2001sm}. Writing the action \eqref{action_730178} as the sum of a free part $\mathcal S_0$ and a perturbation $\mathcal S_1$, and choosing again a gauge in which  $B_{ab} = \frac{1}{3}  H_{abc}\, X^c$, we have
\eq{
  \label{action_pertub_01}
  \mathcal S = \mathcal S_0 + \mathcal S_1 \hspace{40pt}{\rm where}\hspace{40pt}
  \mathcal S_1=\frac{1}{2\pi\alpha'} \:\frac{H_{abc}}{3} \int_\Sigma d^2 z\, X^a
  \partial X^b \,\ov\partial X^c \;.
}
Note that we expect 
${\cal S}_1$ to be a marginal operator, but  not exactly
marginal since it will be obstructed already at second order in $H$.
We will find  evidence for this fact in the following. 

\pagebreak

A correlation function of $N$ operators $\mathcal O_i[X]$ can be computed via the path integral in the usual way 
\eq{
  \label{pi_01}
  \bigl\langle \mathcal O_1 \ldots \mathcal O_N \bigr\rangle
  = \frac{1}{\mathcal Z} \int [dX]
  \,
    \mathcal O_1 \ldots \mathcal O_N \,e^{-\mathcal S[X]} 
  \,,
}
where $\mathcal Z$ denotes the vacuum functional given by $\mathcal Z =\int [dX] \,e^{- S[X]}$. In the limit of small fluxes, it is possible to expand \eqref{pi_01}  in the perturbation $\mathcal S_1$ leading to
\eq{
  \label{master_01}
  \bigl\langle  \mathcal O_1 \ldots \mathcal O_N \bigr\rangle 
  =\quad&\bigl\langle  \mathcal O_1 \ldots \mathcal O_N\bigr\rangle_{0} 
  - \bigl\langle  \mathcal O_1 \ldots \mathcal O_N\: \mathcal S_1 \bigr\rangle_{0} \\[2mm]
  +\:& \frac{1}{2}\, \Bigl[\,
  \bigl\langle  \mathcal O_1 \ldots \mathcal O_N\: \mathcal S^2_1 \bigr\rangle_{0} 
  - \bigl\langle  \mathcal O_1 \ldots \mathcal O_N \bigr\rangle_{0} \times
  \bigl\langle  \mathcal S^2_1 \bigr\rangle_{0} \,\Bigr]
   +    \mathcal O\bigl( H^3\bigr) \;,
}
where for later purpose we have included  corrections up to second order in $H$. The subscript $0$ indicates that the correlator is computed using the free action $\mathcal S_0$, and  we made use of the fact that $\langle \mathcal S_1 \rangle_0$  vanishes. The latter can be verified using the two-point function of two free fields $X^a(z,\ov z)$ (as well as derivatives thereof)
\eq{
  \label{wick01}
  \bigl\langle X^a(z_1,\ov z_1 )\, X^b(z_2,\ov z_2 ) \bigr\rangle_{0}  
  = - \frac{\alpha'}{2} \log |z_1-z_2|^2 \:\delta^{ab}
  \;.
}


\subsection{Three- and two-point function}
\label{sec_basisthree}

In this section,  we determine the correlator of three closed string coordinates $\mathcal X^a(z,\ov z)$ and the correction to the two-point function \eqref{wick01} up to second order in the flux parameter.


\subsubsection*{Three-current correlators}

Let us start by noting that the fields $X^a(z,\ov z)$ appearing in the action \eqref{action_730178} are actually not proper conformal fields of the theory. Only the currents  have a well-defined behavior under conformal transformations. Therefore, as usual, for the free theory we define
\eq{
  \label{def_cur_01}
  J^a(z) = i \hspace{0.5pt}\partial X^a(z) \;, \hspace{60pt}
  \ov J{}^a(\ov z) = i \hspace{0.5pt}\ov \partial X^a(z) \;,
}
which at zeroth order in $H$ are indeed holomorphic and anti-holomorphic, respectively.
The correlator of three say holomorphic currents $J^a(z)$ up to first order in the $H$-flux is then computed using \eqref{master_01} as follows
\eq{
  &\bigl\langle J^a(z_1)\, J^b(z_2)\, J^c(z_3) \bigr\rangle \\
  &\hspace{25pt}= - \bigl\langle J^a(z_1)\, J^b(z_2)\, J^c(z_3) \,\mathcal S_1 \bigr\rangle_0 \\
  &\hspace{25pt}= - \frac{H_{pqr}}{6\pi\alpha'} \int_{\Sigma} d^2 w \,
  \bigl\langle  J^a(z_1)\, J^b(z_2)\, J^c(z_3)  \,
   X^p_0(\omega,\ov\omega ) \,  \partial X^q_0(\omega) \,\ov\partial X^r_0 (\ov \omega)
    \bigr\rangle_0 \;. 
}
The expression in the last line can be evaluated via Wick's theorem employing the two-point function \eqref{wick01} as well as 
\eq{
  \partial_{z_1} \ov \partial_{z_2} \log |z_1-z_2|^2 = -2\pi \, \delta^{(2)}(z_1-z_2) \;.
}
Taking into account the antisymmetry of $H_{abc}$, raising the indices of $H$ with $\delta^{ab}$
and using $z_{ij}=z_i-z_j$, for the  correlators of three currents \eqref{def_cur_01}  (up to first order in the $H$-flux) we find
\eq{
 \bigl\langle J^a(z_1)\, J^b(z_2)\, J^c(z_3) \bigr\rangle  &=  -i\:\frac{{\alpha'}^2}{8}\,H^{abc}\:
   \frac{1}{ z_{12}\, z_{23}\, z_{13}} \;, \\
  \bigl\langle J^a(z_1)\, J^b(z_2)\, \ov J{}^c(\ov z_3) \bigr\rangle &=  -i\:\frac{{\alpha'}^2}{8}\,H^{abc}\:
   \frac{\ov z_{12}}{ z_{12}^2\, \ov z_{23}\, \ov z_{13}} \,, \\
  \bigl\langle \ov J{}^a(\ov z_1)\, \ov J{}^b(\ov z_2)\, J^c(z_3) \bigr\rangle &= +i\:\frac{{\alpha'}^2}{8}\,H^{abc}\:
   \frac{z_{12}}{ \ov z_{12}^2\,  z_{23}\, z_{13}} \;,\\
  \bigl\langle \ov J{}^a(\ov z_1)\, \ov J{}^b(\ov z_2)\, \ov J{}^c(\ov z_3) \bigr\rangle &= +i\:\frac{{\alpha'}^2}{8}\,H^{abc}\:
   \frac{1}{ \ov z_{12}\, \ov z_{23}\, \ov z_{13}} \;.
}

As one can see, these expressions are not purely holomorphic or purely anti-holomorphic, but mixed terms appear. However, we have been using  the currents \eqref{def_cur_01} which are only valid for the free theory. To work at first order in the flux, we should take into account  corrections to \eqref{def_cur_01} linear in $H$. Let us therefore 
define new fields $\mathcal J^a$ and $\ov{\mathcal J}{}^a$ in terms of \eqref{def_cur_01} in the following way
\eq{
  \label{def_cur_04}
  {\cal J}^a (z,\ov z) & = J^a(z)-
  {\textstyle \frac{1}{2}}\hspace{0.5pt}H^a{}_{bc} \, J^b(z) \, 
  {X}^c_R(\ov z)  \;, \\[1mm]
  \ov {\cal J}^a (z,\ov z) & = \ov J{}^a(\ov z)-
  {\textstyle \frac{1}{2}}\hspace{0.5pt}H^a{}_{bc} \, X_L^b( z) \,
  \ov J{}^c(\ov z)  \;.  
}
For the fields \eqref{def_cur_04}, the only non-vanishing correlators of three fields (up to first order in the flux) are then either purely holomorphic or purely anti-holomorphic
\eq{
  \label{cur_cor_19}
 \bigl\langle {\cal J}^a(z_1,\ov z_1)\, {\cal J}^b(z_2,\ov z_2)\, {\cal J}^c(z_3,\ov z_3) \bigr\rangle  
 &=  -i\:\frac{{\alpha'}^2}{8}\,H^{abc}\:
   \frac{1}{ z_{12}\, z_{23}\, z_{13}} \;, \\
  \bigl\langle \ov{\cal J}{}^a(z_1,\ov z_1)\, \ov{\cal J}{}^b(z_2,\ov z_2)\,
  \ov {\cal J}{}^c(z_3,\ov z_3) \bigr\rangle 
  &= +i\:\frac{{\alpha'}^2}{8}\,H^{abc}\:
   \frac{1}{ \ov z_{12}\, \ov z_{23}\, \ov z_{13}} \;.
}
Thus, the corrected fields ${\cal J}^a(z,\ov z)$ and $\ov{\cal J}{}^a(z,\ov z)$, respectively, have 
holomorphic and anti-holomorphic correlation functions. Furthermore, using the equation of motion \eqref{eom_619}, we compute
\eq{
  \ov\partial \mathcal J^a(z,\ov z) = 0 \;, \hspace{60pt}
  \partial \ov{\mathcal J}{}^a(z,\ov z) = 0 \;,
}
so these fields are indeed holomorphic and anti-holomorphic,  and from now on
will be denoted as ${\cal J}^a(z)$ and $\ov{\cal J}{}^a(\ov z)$.
Note also that the correlators \eqref{cur_cor_19}  agree with the three-point function of three
currents in the $SU(2)_k$ WZW-model, 
up to an already anticipated  \cite{Blumenhagen:2010hj}
sign change for the ``structure constants'' $H^{ab}{}_c$.


\subsubsection*{Basic three-point function}

Let us next  define fields ${\cal X}^a(z,\ov z)$ as the integrals of \eqref{def_cur_04}. In particular, we write
\eq{
  \mathcal J^a(z) = i\hspace{0.5pt}\partial \mathcal X^a(z,\ov z) \;,\hspace{60pt}
  \ov{\mathcal J}{}^a(\ov z) = i\hspace{0.5pt}\ov \partial \mathcal X^a(z,\ov z) \;.
}  
The three-point function of three $\mathcal X^a$ up to first order in the $H$-flux can then be obtained by integrating the corresponding  correlators \eqref{cur_cor_19}.
For that purpose, we introduce
the Rogers dilogarithm $L(z)$ which is defined in terms of the usual dilogarithm ${\rm Li}_2(z)$ as follows
\eq{
  L(z)={\rm Li}_2(z) + \frac{1}{ 2} \log (z) \log(1-z)\; .
}
In  appendix \ref{dilog}, we have collected  some useful properties of this function. For the correlator of three fields \eqref{def_x_04} one  obtains\,\footnote{The Rogers dilogarithm also appeared
in the analogous open string three-point function discussed  in \cite{Cornalba:2001sm}.} 
\eq{
  \label{tpf_01}
  &\bigl\langle {\cal X}^a(z_1,\ov z_1)\, {\cal X}^b(z_2,\ov z_2)\,   {\cal X}^{c}(z_3,\ov z_3) \bigr\rangle   
    \\
  &\hspace{80pt}= \frac{{\alpha'}^2}{12}\: H^{abc} 
   \biggl[ L \Bigl( {\textstyle \frac{z_{12}}{z_{13}} } \Bigr) 
  + L \Bigl( {\textstyle \frac{z_{23}}{z_{21}} } \Bigr) 
  + L \Bigl( {\textstyle \frac{z_{13}}{z_{23}} } \Bigr) - {\rm c.c.} \biggr]
  + F \;,
}
where ``c.c.'' stands for complex conjugation and where we have included integration constants collectively denoted by $F$. These have to satisfy
$\partial_{i}\partial_{j} \partial_{k} F=0$
with $i\in \{z_1,\ov z_1\}$, $j\in \{z_2,\ov z_2\}$, $k\in \{z_3,\ov z_3\}$.

Let us note that analogous terms can appear for 
the two-point function of two fields $X^a(z,\ov z)$. Indeed, \eqref{wick01} 
is actually not well-defined  on a two-sphere;  
 only the two-point functions
of the associated currents \eqref{def_cur_01}
are bona-fide conformal objects. In particular, one has the freedom to
add additional terms  $f(z,\ov z)$
satisfying $\partial_{i}\partial_{j} f=0$, where $i\in \{z_1,\ov z_1\}$ 
and $j\in \{z_2,\ov z_2\}$
\eq{
  \label{wick01b}
  \bigl\langle X^a(z_1,\ov z_1 )\, X^b(z_2,\ov z_2 ) \bigr\rangle_{0}  = -\frac{\alpha'}{2}
  \Bigl(\log |z_1-z_2|^2 + f(z_1,\ov z_1) + f(z_2,\ov z_2) \Bigr)\:\delta^{ab}.
}
However, these extra terms  do not contribute  to physical amplitudes and
for this reason we have not included them 
in \eqref{wick01} in the first place.
Furthermore, for the two-point
function on the sphere 
one can  show explicitly that the amplitudes
of vertex operators are indeed independent of the extra terms by  introducing  a background charge.
For  \eqref{tpf_01} it would be interesting to have a
similar proof; here we first proceed with the  assumption that $F=0$.
Later, for the mathematically correct derivation of the four-tachyon amplitude
these terms will become relevant.

To simplify our notation for the following, let us recall from \eqref{setup_01} the relation between the flux $H$ and the flux parameter $\theta$, that is
$\theta^{abc}=\frac{{\alpha'}^2}{12} H^{abc}$, 
and let us introduce
\begin{equation}
   {\cal L}(z)=L(z)+L\left({\textstyle 1-\frac{1}{z}}\right)
    + L\left({\textstyle \frac{1}{1-z}}\right)\; .
\end{equation}
Note that this sum of dilogarithms satisfies the relations
\eq{
\label{sumdilogid}
   {\cal L}(z)={\cal L}\left({\textstyle 1-\frac{1}{z}}\right)
    ={\cal L}\left({\textstyle \frac{1}{1-z}}\right)
    \;,\hspace{40pt}
    {\cal L}(z)+{\cal L}(1-z)=3\hspace{0.5pt} L(1)={\textstyle  \frac{\pi^2}{2}}\; .
}
The correlation function \eqref{tpf_01} of three fields $\mathcal X^a(z,\ov z)$ in the $H$-flux background can then be written as
\begin{equation}
  \label{three-point_01}  
  \bigl\langle {\cal X}^a(z_1,\ov z_1)\, {\cal X}^b(z_2,\ov z_2)\,    {\cal X}^{c}(z_3,\ov z_3) \bigr\rangle 
  = {\theta^{abc}} \Bigl[{\cal L}
   \bigl( {\textstyle \frac{z_{12}}{ z_{13}} }\bigr) - {\cal L}
   \bigl({\textstyle \frac{\ov z_{12}}{ \ov z_{13}}}\bigr) \Bigr]\; .
\end{equation}


\subsubsection*{Correction to the two-point function}

In our above discussion, we  have seen  that at linear order in 
the $H$-flux, the three-point function for the corrected fields $\mathcal X^a$ is  conformally
invariant. However, at second order this is no longer true which we want to illustrate for the two-point function.

For the regular fields $X^a(z,\ov z)$ the two-point function up to second order in the flux can be computed using formula \eqref{master_01}. Since the correction at linear order in $H$ vanishes for a correlator of two fields, we are left with expressions quadratic in the perturbation $\mathcal S_1$
\eq{
  \delta_2 \bigl\langle  X^a(z_1,\ov z_1) X^b(z_2,\ov z_2)  \bigr\rangle
  =&+ \frac{1}{2}\,
  \bigl\langle  X^a(z_1,\ov z_1) X^b(z_2,\ov z_2)\: \mathcal S^2_1 \bigr\rangle_{0} \\
  &- \frac{1}{2}\,
   \bigl\langle  X^a(z_1,\ov z_1) X^b(z_2,\ov z_2) \bigr\rangle_{0} \times
  \bigl\langle  \mathcal S^2_1 \bigr\rangle_{0} \;.
}
Recalling then  formula \eqref{action_pertub_01} for $\mathcal S_1$, this correction reads
\eq{
  \label{tpf_corr_78}
  \hspace{185pt}&\hspace{-185pt}
  \delta_2 \bigl\langle  X^a(z_1,\ov z_1) X^b(z_2,\ov z_2)  \bigr\rangle
   = \frac{1}{2\,(6\pi \alpha')^2} \: H_{mno} \, H_{pqr} 
   \int_{\Sigma} d^2 w_1  \int_{\Sigma} d^2 w_2  \\[0.1cm]
 \bigl\langle :\! X^a(z_1,\ov z_1) X^b(z_2,\ov z_2)\!:   
 &:\! X^m(w_1,\ov w_1)\, \partial X^n(w_1) \,\ov\partial X^o(\ov w_1) \!:  \\
 &:\! X^p(w_2,\ov w_2)\hspace{4.5pt} \partial X^q(w_2) \hspace{3.5pt}
 \ov\partial X^r(\ov w_2) \!: \bigr\rangle_{0} \,.
}
The expression in the last two lines can be evaluated using Wick contractions but, as explained in more detail in appendix \ref{app_twopt}, the integrals in \eqref{tpf_corr_78} have to be regularized.
This can be done by removing a small  disc specified by $|w_1-w_2|<\epsilon$ for $\epsilon\ll1$
from the integration region, which defines an ultra-violet cutoff.
The corresponding computation is shown in appendix \ref{app_twopt} and the result reads
\eq{
   \delta_2 \bigl\langle X^a(z_1,\ov z_1) \, X^b(z_2,\ov z_2)  \bigr\rangle =
   \frac{{\alpha'}^2}{8} H^a{}_{pq}\, H^{bpq} \, \log |z_1-z_2|^2\;  \log\epsilon \;.
}
Therefore, the perturbation ${\cal S}_1$ ceases to be marginal at second order in the flux and the theory is not conformally invariant. 
Writing finally the two-point function as
$\langle X^a(z_1,\ov z_1) X^b(z_2,\ov z_2)  \rangle= -\frac{\alpha'}{2}  \log |z_1-z_2|^2\hspace{0.5pt} g^{ab}$, we find a renormalization group flow equation for the inverse world-sheet metric $g^{ab}$ of the form 
\eq{
  \mu\: \frac{\partial\hspace{0.5pt} g^{ab}}{ \partial \mu} =-\frac{\alpha'}{4} 
   H^a{}_{pq}\, H^{bpq}  \; ,
}
which agrees with equation \eqref{targeteom} for constant space-time metric,  $H$-flux and dilaton.


\subsection{Structure of CFT$_H$}

In equation \eqref{cur_cor_19} we have seen that the redefined fields ${\cal J}^a(z)$ feature
a holomorphic three-point function up to first order in the flux. One may therefore suspect
that up to linear order in $H$ there exists a conformal field theory, in the following denoted by CFT$_H$. 
In this subsection we  provide more evidence for
this observation by analyzing the operator product expansion (OPE) for the
currents and the energy-momentum tensor. Our final goal
is to define vertex operators for CFT$_H$ which will allow 
us to compute string scattering amplitudes.


\subsubsection*{Current algebra and energy-momentum tensor}

Let us first study  the fields ${\cal J}^a(z)$ and $\ov{\mathcal J}{}^a(\ov z)$ defined in \eqref{def_cur_04} in more detail. Their non-vanishing two-point function up to first order in  $H$ is readily found to be 
\eq{
  \label{ope_98}
  \arraycolsep2pt
  \begin{array}{ccccc}
  \displaystyle \bigl\langle {\cal J}^a (z_1) {\cal J}^b (z_2) \bigr\rangle 
  &=& 
  \displaystyle \bigl\langle J^a(z_1)\, J^b(z_2) \bigr\rangle_0 
  &=&
  \displaystyle  \frac{\alpha'}{2} \:\frac{1}{(z_1-z_2)^2}\:\delta^{ab} \;, \\[3mm]
  \displaystyle \bigl\langle \ov{\cal J}{}^a (\ov z_1)\, \ov{\cal J}{}^b (\ov z_2) \bigr\rangle 
  &=& 
  \displaystyle \bigl\langle \ov J{}^a(z_1)\, \ov J{}^b(z_2) \bigr\rangle_0 
  &=& 
  \displaystyle \frac{\alpha'}{2} \:\frac{1}{(\ov z_1-\ov z_2)^2}\:\delta^{ab} \;, 
  \end{array}
}
where we employed the definition \eqref{def_cur_01} of the currents $J^a(z)$
as well as the two-point function of the fields $X^a(z,\ov z)$ shown in
\eqref{wick01}. This result reflects that, even though the coordinates $X^a$ are
corrected to $\mathcal X^a$,  at linear order in the flux the metric
is not. Indeed, in the last subsection we have seen that the metric
receives correction only at second order in $H$.
Taking then into account the three-point functions  \eqref{cur_cor_19} of the fields $\mathcal J^a(z)$ and $\ov{\mathcal J}{}^a(\ov z)$, with the help of \eqref{ope_98} we can construct the following OPEs 
\eq{
  \label{OPE_01}
  \arraycolsep2pt
  \begin{array}{cclclcl}
  \displaystyle {\cal J}^a(z_1)\; {\cal J}^b(z_2) 
  &=&
  \displaystyle  \frac{\alpha'}{2} \,\frac{\delta^{ab}}{(z_1-z_2)^2}
  &-&
  \displaystyle \frac{\alpha'}{4}\, \frac{i\, H^{ab}{}_c}{z_1-z_2}\: {\cal J}^c(z_2) 
  &+&
  \displaystyle  {\rm reg.} \;, \\[4mm] 
  \displaystyle \ov{\cal J}{}^a(\ov z_1)\; \ov{\cal J}{}^b(\ov z_2) 
  &=& 
  \displaystyle \frac{\alpha'}{2} \,\frac{\delta^{ab}} {(\ov z_1-\ov z_2)^2}
  &+&
  \displaystyle \frac{\alpha'}{4}\, \frac{i\, H^{ab}{}_c}{\ov z_1-\ov z_2}\: \ov{\cal
    J}{}^c(\ov z_2) 
  &+&
  \displaystyle {\rm reg.}\;,
  \end{array}
}
where ``reg.'' stands for regular terms and where the OPEs 
between ${\cal J}^a(z)$ and $\ov{\cal J}{}^b(\ov z)$ are purely regular.
Note that \eqref{OPE_01} defines two independent non-abelian current algebras 
with structure constants $f^{ab}{}_c\simeq H^{ab}{}_c$. The only difference to the usual 
expressions is an opposite relative sign for $H^{ab}{}_c$ between the
holomorphic and anti-holomorphic parts.

Next, we analyze the  energy-momentum tensor. Since the Kalb-Ramond
part of the world-sheet action \eqref{action_730178}  is independent of the
world-sheet metric, we obtain the result for the free theory which reads
$T(z)= \frac{1}{\alpha'}\: \delta_{ab}:\! {J}^a {J}^b\!:\!(z)$
and $\ov T(\ov z) = \frac{1}{\alpha'}\: \delta_{ab} :\! \ov{J}{}^a\ov{J}{}^b\!:\!(\ov z)$.
However, due to the antisymmetry of $H_{abc}$, up to linear order in the flux we can  write
\eq{
  \label{def_emt_01}
  {\cal T}(z) = \frac{1}{\alpha'}\: \delta_{ab}:\! {\cal J}^a {\cal J}^b\!:\! (z) \;, \hspace{40pt}
  \ov{\cal T}(\ov z) = \frac{1}{\alpha'}\: \delta_{ab} :\! \ov{\cal J}{}^a\ov{\cal J}{}^b\!:\! (\ov z) \;.
}
The antisymmetry of $H$ furthermore implies that the OPEs of two energy-mo\-men\-tum tensors take the form
\eq{
  \label{OPE_02}
  \arraycolsep2pt
  \begin{array}{ccccc}
  \displaystyle {\cal T}(z_1)\; {\cal T}(z_2) 
  &=&
  \displaystyle  \frac{{c/2}}{(z_1-z_2)^4} + \frac{2\, {\cal T}(z_2)}{(z_1-z_2)^2}
  + \frac{\partial {\cal T}(z_2)}{z_1-z_2} 
  &+& {\rm reg.} \;, \\[4mm]
  \displaystyle \ov{\cal T}(\ov z_1)\; \ov {\cal T}(\ov z_2) 
  &=&
  \displaystyle  \frac{c/2}{(\ov z_1-\ov
   z_2)^4} + \frac{2\, \ov{\cal T}(z_2)}{(\ov z_1-\ov z_2)^2}
  + \frac{\partial \ov{\cal T}(z_2)}{\ov z_1-\ov z_2} 
  &+& {\rm reg.} \;, 
  \end{array}
}
with $ {\cal T}(z_1) \, \ov {\cal T}(\ov z_2)$ regular.
We therefore find two copies of the  Virasoro algebra with the same central
charge $c$ as for the free theory.
(In the present case of a flat three-dimensional space, this means $c=3$.)
Moreover, using \eqref{OPE_01} and again the antisymmetry of $H$, one
straightforwardly finds 
\eq{
  \label{OPE_03}
  &{\cal T}(z_1)\, {\cal J}^a(z_2) 
  =  \frac{{\cal J}^a(z_2)}{(z_1-z_2)^2} +
     \frac{\partial {\cal J}^a(z_2)}{z_1-z_2}+{\rm reg.} \;,\\[2mm]
  &\ov{\cal T}(\ov z_1)\, {\cal J}^a(z_2) 
  = {\rm reg.} \;,
}
and analogously for the anti-holomorphic parts.
The fields ${\cal J}^a(z)$ and $\ov{\cal J}{}^a(\ov
z)$ are therefore primary fields of  conformal dimension $(1,0)$ and $(0,1)$ with respect to $\mathcal T(z)$ and $\ov{\mathcal T}(\ov z)$; and in  CFT$_H$ they are indeed non-abelian  currents.


\subsubsection*{Vertex operator for the tachyon}

Let us now define vertex operators. In analogy to the free theory of  a closed string without $H$-flux, which in a compact space can  have momentum $p_a$ and winding $w^b$, we define  left- and right-moving momenta $k_{L/R}$ as 
\eq{
  \label{def_mom_76}
  k_{L}^a = p^a + \frac{w^a}{\alpha'} \;, \hspace{60pt}
  k_{R}^a = p^a - \frac{w^a}{\alpha'} \;.
}
The vertex operator for the perturbed theory should then be written in the following way
\eq{
  \label{def_vo7023}
  {\cal V}(z,\ov z) = \, :\!\exp \bigl( i\hspace{0.5pt} k_L\cdot {\cal X}_L +
  i\hspace{0.5pt} k_R \cdot {\cal X}_R \bigr) \!: \;,
}
where we again employ the short-hand notation $k_L\cdot {\cal X}_L= k_{La} {\cal X}^a_L$. The left- and right-moving fields ${\cal X}^a_{L/R}$ are obtained via
integration of  the currents, and their OPEs can be computed from \eqref{OPE_01} as
\eq{
  \label{OPE_04}
  \arraycolsep2pt
  \begin{array}{lclclcl}
  \displaystyle {\cal J}^a(z_1)\, {\cal X}^b_L(z_2) 
  &=&
  \displaystyle  -i\:\frac{\alpha'}{2} \:\frac{\delta^{ab}}{z_1-z_2}
  &+&
  \displaystyle  \frac{\alpha'}{4}\,  H^{ab}{}_c\; {\cal J}^c(z_2)\, \log(z_1-z_2) 
  &+&{\rm reg.} \;, \\[3mm]
  \displaystyle \ov{\cal J}{}^a(\ov z_1)\, {\cal X}^b_R(\ov z_2) 
  &=&
  \displaystyle  -i\:\frac{\alpha'}{2} \:\frac{\delta^{ab}}{\ov z_1-\ov z_2}
  &-&
  \displaystyle  \frac{\alpha'}{4}\, H^{ab}{}_c\; \ov{\cal  J}{}^c(\ov z_2)\, \log(\ov z_1-\ov z_2) 
  &+& {\rm reg.} \;.
  \end{array}
}
Now, recall that in the free theory the tachyon vertex operator
is a primary field  of conformal dimension $(h,\ov h)=(\frac{\alpha'}{4}\,
k_L^2,\frac{\alpha'}{4}\, k_R^2)$, and
in covariant quantization of the bosonic string
physical states are given by primary fields of conformal
dimension $(h,\ov h)=(1,1)$. 
In the deformed theory CFT$_H$, we also require that vertex operators ${\cal V}(z,\ov z)$ 
are primary with respect to ${\cal T}(z)$ and 
$\ov {\cal T}(\ov z)$ which, due to the linear  corrections  \eqref{OPE_04},
is not guaranteed a priori.
However, it is again the antisymmetry of $H$ which implies 
\eq{
   \arraycolsep2pt
   \begin{array}{lclclcl}
   \displaystyle {\cal T}(z_1)\, {\cal V}(z_2,\ov z_2) 
   &=& 
   \displaystyle \frac{1}{(z_1-z_2)^2} \, \frac{\alpha' k_L\cdot k_L}{4} \,   {\cal V}(z_2,\ov z_2)
   &+&
   \displaystyle \frac{1}{z_1-z_2} \, \partial {\cal V}(z_2,\ov z_2) 
   &+&{\rm reg.} \;, \\[3mm]
   \displaystyle \ov{\cal T}(\ov z_1)\, {\cal V}(z_2,\ov z_2) 
   &=& 
   \displaystyle \frac{1}{(\ov z_1-\ov z_2)^2} \, \frac{\alpha' k_R\cdot k_R}{4} \,  {\cal V}(z_2,\ov z_2)
   &+&
   \displaystyle \frac{1}{\ov z_1-\ov z_2} \, \ov\partial {\cal V}(z_2,\ov z_2) 
   &+&{\rm reg.} \;.
   \end{array}  
}
Thus,  the vertex operator \eqref{def_vo7023} is  primary and 
has conformal dimension $(h,\ov h)=(\frac{\alpha'}{4}\,
k_L^2,\frac{\alpha'}{4}\, k_R^2)=(1,1)$. It is therefore a physical quantum
state of the  deformed theory.
Classically, this corresponds to the 
fact  that $\mathsf{X}_1^a(z,\ov z)$, as defined in \eqref{def_x_04},
solves the classical sigma-model equation of motion \eqref{eom_619} 
up to linear order in $H$,
with  the zeroth order classical tachyonic solution given by
\eq{
\label{classtac}
   \mathsf X_0^a(\sigma,\tau)=x^a + \alpha' p^a \, \tau + w^a\, \sigma\;.
}


\subsubsection*{Momentum and winding for the tachyon vertex operator}

We continue our discussion and note that in the free theory without flux, the usual vertex operator
$V(z,\ov z) = :\!\exp ( k_L\cdot {X}_L + k_R \cdot {X}_R )\!:$
 carries center of mass momentum $p^a$ and winding $\omega_a$.
We want to determine the analogue for the perturbed theory by computing the following OPE in CFT$_H$
\eq{
  \label{some_ope_739}
   {\cal J}^a(z_1)\, {\cal V}(z_2,\ov z_2) = &\quad
   \frac{1}{z_1-z_2}  \frac{\alpha' k^a_L}{2}\,   {\cal V}(z_2,\ov z_2)\\
   &+i\,\frac{\alpha'}{4} \log (z_1-z_2) \,  H^{a}{}_{bc}\; k^b_L :\! {\cal  J}{}^c\, {\cal
     V}\!:\!(z_2,\ov z_2) + {\rm reg.} \;,
}
and similarly  for the anti-holomorphic part. 
Note that in \eqref{some_ope_739} a logarithmic term appears, which 
is also true for more general vertex operators to be discussed below.
There are two possibilities to deal with this term:
\begin{enumerate}

\item In order to have a well-defined conventional CFT, such terms must be absent implying the constraint $H^{a}{}_{bc}\, k^b_L=0$. The momenta are therefore  forced to be transversal to the $H$-flux which  would trivialize most of the
results obtained in the following.

\item The second possibility is that generically CFT$_H$  is a logarithmic CFT (LCFT) in which
such terms have a physical meaning.

\end{enumerate}
In this paper, we take the latter point of view so that the logarithmic terms should not be eliminated from the very beginning, but should be treated
as carrying  vanishing conformal dimension.
More concretely, we may add a term of the form $\log (\ov z_1-\ov z_2) \,  H^{a}{}_{bc}\; k^b_L :\! {\cal  J}{}^c\, {\cal V}\!:\!(z_2,\ov z_2)$ to the OPE \eqref{some_ope_739},  which is regular in the holomorphic variable $z_1$. Therefore, in the above OPE the logarithm can be replaced by $\log|z_1-z_2|^2$ implying also that  \eqref{some_ope_739} is single valued.

Let us continue and consider the zero mode ${\cal P}_L^a$ of ${\cal J}^a(z)$, 
which can be defined via a contour integral. To determine the $\mathcal P_L^a$ eigenvalue of the vertex operator we compute
\eq{
  \label{momencom}
     \lim_{z_2,\ov z_2 \to 0} {\cal P}_L^a\, {\cal V}(z_2,\ov z_2)\bigr|0\bigr\rangle
     &=
     \lim_{z_2,\ov z_2\to 0} \oint \frac{dz_1}{2\pi i} \, {\cal J}^a(z_1)\, {\cal V}(z_2,\ov z_2)
     \bigr|0\bigr\rangle \\
     &= \frac{\alpha' \hspace{0.5pt}k_L^a}{2} \lim_{z_2,\ov z_2\to 0} {\cal V}(z_2,\ov z_2)
     \bigr|0\bigr\rangle \;.
}
Since $\mathcal P^a_L$  has 
a contribution from the Kalb-Ramond part of the sigma-model, 
it is equivalent to the canonical momentum (though it differs by a numerical prefactor).
But, similar to the situation for the energy-momentum tensor,
the physical momentum should be related to the uncorrected expression, that is
$P_L^a=\oint \frac{dz}{2\pi i} J^a(z)$. Therefore, using \eqref{def_cur_04} we compute
in the perturbed background (up to first order in $H$)
\eq{
  \label{momencomrealb}
   &\lim_{z_2,\ov z_2\to 0} {P}_L^a\, {\cal V}(z_2,\ov z_2) \bigr|0\bigr\rangle \\
   =&\lim_{z_2,\ov z_2\to 0} \oint \frac{dz_1}{2\pi i}\,
   J^a(z_1)\, {\cal V}(z_2,\ov z_2)\bigr|0\bigr\rangle   \\
   =& \lim_{z_2,\ov z_2\to 0} \oint \frac{dz_1}{2\pi i}
   \left[ {\cal J}^a(z_1)\, {\cal V}(z_2,\ov z_2)
    +\frac{1}{2}\, H^a{}_{bc} \, J^b(z_1) \, X^c_R(\ov z_1)\, {V}(z_2,\ov z_2)\right]
       \bigr|0\bigr\rangle \;.
}
Note that the first term in the last line is \eqref{momencom}, and since the second term is already
linear in the flux we can work with the free theory.
In particular, the second term can be evaluated to be proportional to 
$ H^a{}_{bc}\,  k_L^b\, k_R^c$, so in order for 
the tachyon vertex operator in CFT$_H$ to carry momenta $(k_L,k_R)$ we have to require
\eq{
  \label{M1860}
  0= H^a{}_{bc}\, k_L^b\, k_R^c \simeq
  H^a{}_{bc}\, p^b\, w^c  \simeq  \bigl[\, \vec p \times \vec w \,\bigr]^a\;,
}
where we have used that $H^a{}_{bc}\simeq \epsilon^a{}_{bc}$.
Again, this constraint has  a corresponding classical analogue. Indeed, 
taking the classical tachyon \eqref{classtac} and requiring
\eq{
   \frac{1}{ 2\pi} \int_0^{2\pi} d\sigma\, \partial_\tau {\cal X}^a = \alpha'\, p^a \;, 
   \hspace{40pt}
   \frac{1}{2\pi} \int_0^{2\pi} d\sigma\, \partial_\sigma {\cal X}^a =  \omega^a \;,
}  
yields again the constraint \eqref{M1860}.\footnote{For $\vec p \times \vec
  w\ne\vec 0$ the physical momenta are not conserved, which is the
higher-dimensional analogue of the well-known cyclotron orbits
 arising for point particles
moving in a constant magnetic field.} Intriguingly, it
can also be derived by requiring that the vertex operator 
 of the free-theory $V(z,\ov z)$ is a primary field of the perturbed one. 
As can readily  be shown, classically this means that 
the free tachyon solution
\eqref{classtac} remains a solution of the $H$-corrected
equation of motion when \eqref{M1860} is satisfied.
We will analyze the consequences of this constraint further
in section \ref{sec_Tdual}.


\subsubsection*{Vertex operator for the graviton}

Let us also  consider the vertex operator for the ``graviton'' in the perturbed theory CFT$_H$, which we define as
\eq{
  \label{gravvert}
  {\cal V}_G(z,\ov z) 
  = \zeta_{ab}   :\! {\cal J}^a \, \ov{\cal J}{}^b  \exp \bigl( i\hspace{0.5pt} k_L\cdot {\cal X}_L +
  i\hspace{0.5pt} k_R \cdot {\cal X}_R \bigr) \!:  .
}
The OPE with the holomorphic energy-momentum tensor can be computed using
\eqref{OPE_04}. Employing the antisymmetry of $H_{abc}$, one finds
\begin{align}
  \nonumber
   {\cal T}(z_1)\, {\cal V}_G(z_2,\ov z_2) =&\quad
    \frac{\alpha'}{2} \frac{ \zeta_{ab}\,   k_L^a }{(z_1-z_2)^3} 
    :\!\ov{\cal J}{}^b\exp(i\hspace{0.5pt}k\cdot {\cal X})\!: \\
  \label{OPE_TG}
   &+ \left[\frac{\frac{\alpha'}{4} k_L^2+1}{(z_1-z_2)^2} +
   \, \frac{\partial} {z_1-z_2} \right]  {\cal V}_G(z_2,\ov z_2)
   \\
   \nonumber
   & + \frac{i\hspace{0.5pt} \alpha'}{4}\: \frac{1+\log(z_1-z_2)}{(z_1-z_2)^2} \:    
     H^a{}_{cd}\, k_L^c\, \zeta_{ab}   :\! {\cal J}^d \, \ov{\cal J}{}^b
     \exp(i\hspace{0.5pt}k\cdot {\cal X})\!: \;,
\end{align}
and similarly for the anti-holomorphic part. 
Requiring that ${\cal V}_G$ is a primary field of conformal
dimension $(1,1)$, the first two lines in \eqref{OPE_TG} give the usual 
on-shell conditions
$\zeta_{ab}\, p^a = \zeta_{ab}\, \omega^a=0$ and $k_L^2=k_R^2=0$.

Note that  we again obtain a logarithmic term in the OPE, which we could
either require to be absent from the very beginning or interpret as an indication for a
logarithmic CFT. In the first case, the last line in \eqref{OPE_TG} (and its anti-holomorphic counterpart) implies the  additional transversality conditions
\eq{
  \label{gravbsigma}
  H^{abc} p_b\, \zeta_{cd}=H^{abc} p_b\, \zeta_{dc}=0 \;, \hspace{40pt}
  H^{abc} w_b\, \zeta_{cd}=H^{abc} w_b\, \zeta_{dc}=0 \; .
} 
In the second case, tolerating the $\log$-term, we find that
the energy-momentum tensor does not act diagonally. 
For instance, in the case of a three-torus $\mathbb T^3$ one can
see that  $L_0=1$ leads to the mass eigenvalues
\eq{
   \label{masssplit}
   m_L^2= 0\;,
   \hspace{50pt}
    m_L^2=\pm \theta \left( \sum_{a=1}^3
     (k^a_L)^2\right)^{\frac 12} \;,
}
where $\theta$ is given by the flux parameter $\theta_{abc}=\theta \epsilon_{abc}$
and the sum runs only over the momenta longitudinal to the
$\mathbb T^3$.
Thus, some of the former massless states become
massive and in particular tachyonic, though 
level matching will  eliminate some of these states.
Therefore, we obtain the physically acceptable result
that some of the longitudinal fluctuations around the constant $H$-flux
background become massive.
It would be interesting to completely determine the mass spectrum
for  this LCFT, but here (in particular in section \ref{sec_fluxedVS}) we take a different approach and
identify the new mass spectrum of the theory via
the poles of the four-tachyon scattering amplitude.


\subsection{T-duality}
\label{sec_Tdual}

As expected from the string equations of motion, in the last subsection we 
have found a bona-fide conformal field theory CFT$_H$, which
describes the sigma-model for a flat metric and constant $H$-flux up to
linear order.
However, we are also interested in backgrounds T-dual to the $H$-flux configuration.

On the level of the CFT, T-duality is usually realized as
a reflection on the right-moving coordinates. Since 
the corrected fields ${\cal X}^a(z,\ov z)$  still admit
a split into a holomorphic and an anti-holomorphic piece,
we define T-duality on the  world-sheet action along direction $\mathcal X^a$ as 
\eq{
  \begin{array}{c}
  {\cal X}_L^a(z) \\[1mm]
  {\cal  X}_R^a(\ov z)   
  \end{array}
  \qquad
  \xrightarrow{\;\mbox{\scriptsize T-duality}\;}
  \qquad
  \begin{array}{c}
  +{\cal X}_L^a(z)\;, \\[1mm]
  -{\cal  X}_R^a(\ov z) \;.
  \end{array}
}
Clearly, for the currents this implies
\eq{
  \begin{array}{c}
  {\cal J}^a(z) \\[1mm]
  \ov{\cal  J}{}^a(\ov z)   
  \end{array}
  \qquad
  \xrightarrow{\;\mbox{\scriptsize T-duality}\;}
  \qquad
  \begin{array}{c}
  +{\cal J}^a(z)\;, \\[1mm]
  -\ov{\cal  J}{}^a(\ov z) \;,
  \end{array}
}
and so  the ``structure constants'' $H^{ab}{}_c$ 
in the  the anti-holomorphic OPE \eqref{OPE_01} receive an additional minus sign
when performing a T-duality transformation.

In the next section, we compute scattering amplitudes
for tachyon vertex operators in the $H$-flux background.
There we allow for both momentum and winding along the 
directions of our three-dimensional (compact) space specified by \eqref{setup_01}.
From table \ref{tableTdual} we infer that these
scattering amplitudes  in the $H$-flux background are related to 
the scattering of appropriate momentum and winding states in the
$\omega$-, $Q$- and $R$-flux backgrounds. However, in the T-dual
models, we are particularly interested in pure momentum
scattering, as from there one would derive the
low-energy effective action as a (ordinary) derivative expansion.
Now, in the previous section we have seen that a tachyon vertex operator $\mathcal V(z,\ov z)$
indeed corresponds to a physical state, but its
usual momentum and winding quantum numbers are only uncorrected
if $ \vec p \times \vec w =\vec 0$.
Recall that classically this means that the free tachyon solution
of the sigma-model equations of motion remains to be a solution
of the $H$-corrected ones. Thus, we expect that 
after imposing the above constraint,
any effect we derive at linear order in $H$ cannot be a consequence of 
the linear redefinition of the classical solution of the tachyon,
but reflects a property of the uncorrected solution.

In table \ref{tablemomwind} we have made the T-duality relations  more explicit, 
which we explain in some detail.
\begin{table}[t]
\centering
\renewcommand{\arraystretch}{1.2}
\begin{tabular}{|cc|cc|cc|cc|}
\hline
\multicolumn{2}{|c}{$H$-flux} & 
\multicolumn{2}{|c}{$\omega$-flux} & 
\multicolumn{2}{|c}{$Q$-flux} &  
\multicolumn{2}{|c|}{$R$-flux} 
\\ \hline\hline
$\langle p_1,p_2,p_3\rangle^-$ & \checkmark &  
$\langle p_1,p_2,w_3\rangle^-$ & \checkmark & 
$\langle p_1,w_2,w_3\rangle^-$ & \checkmark &  
$\langle w_1,w_2,w_3\rangle^-$& \checkmark 
\\ \hline
$\langle p_1,p_2,w_3\rangle^+$ & $\times$ & 
$\langle p_1,p_2,p_3\rangle^+$ & $\times$ & 
$\langle p_1,w_2,p_3\rangle^+$ & $\times$ & 
$\langle w_1,w_2,p_3\rangle^+$ & $\times$ 
\\ \hline
$\langle p_1,w_2,w_3\rangle^-$ &  $\times$ & 
$\langle p_1,w_2,p_3\rangle^-$ & $\times$ & 
$\langle p_1,p_2,p_3\rangle^-$ & $\times$ & 
$\langle w_1,p_2,p_3\rangle^-$ & $\times$ 
\\ \hline
$\langle w_1,w_2,w_3\rangle^+$ &  \checkmark &
$\langle w_1,w_2,p_3\rangle^+$ & \checkmark &
$\langle w_1,p_2,p_3\rangle^+$ & \checkmark &
$\langle p_1,p_2,p_3\rangle^+$ & \checkmark 
\\ \hline
\end{tabular}
\caption{\small  T-duality relations of momentum and winding 
mode scattering  in the four three-form flux backgrounds. 
Entries in the same row are related via the T-dualities from
table \ref{tableTdual}, and the upper index indicates the relative sign between the  holomorphic
and anti-holomorphic part in the three-point function  $\langle \mathcal X^a \mathcal X^b \mathcal X^c\rangle$. 
The symbol thereafter indicates whether 
the condition \eqref{M1860} is satisfied.\label{tablemomwind}} 
\end{table}
\begin{itemize}

\item From the first row we infer that the effective field theory 
(with space-time derivatives $\frac{\partial}{\partial X^a}$) 
for tachyons in the $H$-flux
background is expected to be reliably computable (in the sense
explained above) 
from scattering amplitudes of pure momentum tachyons, since in this case
\eqref{M1860} is satisfied. The 
basic three-point function of the 
coordinates is then given by \eqref{three-point_01}.
Therefore, from now on we denote
\eq{ 
 \bigl\langle {\cal X}^a(z_1,\ov z_1)\, {\cal X}^b(z_2,\ov z_2)\, {\cal X}^c(z_3,\ov z_3)\bigr\rangle^{-}\overset{\rm def}{=}
    {\theta^{abc}} \Bigl[{\cal L}
   \bigl( {\textstyle \frac{z_{12}}{ z_{13}} }\bigr) - {\cal L}
   \bigl({\textstyle \frac{\ov z_{12}}{ \ov z_{13}}}\bigr) \Bigr]\; .
}

\item The next row shows that a scattering of pure momentum  modes
in the geometric flux background is related
to the scattering of $(p_1,p_2,w_3)$  modes in the $H$-flux
background  by T-duality. However, in this case ${\vec w}\times {\vec p}\ne \vec0$
and we cannot exclude that any effect we derive
at linear order in $H$ is  just reflecting
the linear redefinition \eqref{def_x_04} of the space coordinate.
The same situation occurs for pure momentum scattering in the
 $Q$-flux background.

\item The last row in the table shows that only for the case
of $R$-flux we can again reliably compute the scattering
amplitudes for pure momentum tachyons. By T-duality, they  are   related
to the scattering of pure winding states in
the $H$-flux background. Employing T-duality for  this $R$-flux background, 
the basic three-point function for pure momentum states reads
\eq{ 
\bigl\langle {\cal X}^a(z_1,\ov z_1)\, {\cal X}^b(z_2,\ov z_2)\, {\cal X}^c(z_3,\ov z_3)\bigr\rangle^{+}\overset{\rm def}{=}
{\theta^{abc}} \Bigl[{\cal L}
   \bigl( {\textstyle \frac{z_{12}}{ z_{13}} }\bigr) + {\cal L}
   \bigl({\textstyle \frac{\ov z_{12}}{ \ov z_{13}}}\bigr) \Bigr]\; .
}

\end{itemize}

These considerations are also consistent with the results in \cite{Lust:2010iy} from which
 the commutation relations between coordinates $X^a$ and their duals  $\tilde X^a$  can be obtained.\footnote{More precisely, the $\omega$-, $Q$- and $R$-flux cases with elliptic monodromies were explicitly discussed in  \cite{Lust:2010iy}; the commutation relations for parabolic fluxes will be discussed in \cite{ALLP}.} 
Furthermore, using the canonical commutation relations $[X^1,p_1]={\rm const.}$ and 
$[\tilde X^1,w_1]={\rm const.}$, one can derive three-brackets for the four different cases.
These results are summarized in table \ref{tablemomwind2}.

\begin{table}[t]
\centering
\renewcommand{\arraystretch}{1.3}
\tabcolsep10pt
\begin{tabular}{|c||c|c|}
\hline
Flux  & Commutators & Three-brackets \\  \hline\hline
$H$-flux & $[\tilde X^2,\tilde X^3]\simeq w_1$ & $[\tilde X^2,\tilde X^3, \tilde X^1]$\\
$\omega$-flux & $[\tilde X^2,X^3]\simeq w_1$ & $[\tilde X^2,X^3, \tilde X^1]$\\
$Q$-flux & $[ X^2,X^3]\simeq w_1$ & $[X^2,X^3, \tilde X^1]$ \\
$R$-flux & $[ X^2,X^3]\simeq p_1$ & $[ X^2,X^3, X^1]$ \\ \hline
\end{tabular}
\caption{\small  Non-vanishing commutators and three-brackets in the four flux backgrounds.
\label{tablemomwind2} } 
\end{table}


\section{Tachyon scattering amplitudes}
\label{sec_threep}

In this section, by computing higher $N$-point scattering amplitudes 
of tachyon vertex operators and discussing their pole structure, 
we want to infer properties of the theory.
This is in the same spirit as for the famous Veneziano and 
Virasoro-Shapiro four-point amplitudes, which
contain information about the Regge resonances
as well as about the three-point couplings involving two tachyons.


\subsection{Three-tachyon amplitude}
\label{sec_threetac}

We start with the  three-tachyon  amplitude.
As discussed above, we focus on a compact three-dimensional space and therein
we are interested in pure momentum $(p_1,p_2,p_3)$ or pure winding $(w_1,w_2,w_3)$ state scattering, where the latter is related by three T-dualities to pure momentum scattering in the $R$-flux
background. We therefore consider vertex operators of form
\eq{
  \label{vo_tach_19}
  \arraycolsep2pt
  \begin{array}{lclcl}
  {\cal V}_i^- &\equiv& 
  {\cal V}_{p_i}(z_i,\ov z_i) &=& 
  \displaystyle :\! \exp \bigr(\, i\hspace{0.5pt} p_i \cdot {\cal X}(z_i,\ov z_i) \bigl) : \;, \\[1.75mm]
  {\cal V}_i^+ &\equiv& 
  {\cal V}_{w_i}(z_i,\ov z_i) &=& 
  \displaystyle :\! \exp \bigr(\, i\hspace{0.5pt} w_i \cdot \widetilde{\cal X}(z_i,\ov z_i) \bigl) : \;,
  \end{array}
}
where $\widetilde{\cal X}={\cal X}_L-{\cal X}_R$.  Note that here and in the following
we  employ the short hand notation ${\cal V}^\mp_i$ and $V^\mp_i$
for the vertex operators of the perturbed and free theory, respectively.
Furthermore, since we can consider  ${\cal V}_i^+$ as the momentum vertex operator
in the  $R$-flux background, in the following we set $w|_H\to p|_R$.
However, in string theory one has to work with vertex operators integrated over the world-sheet. We therefore define 
\eq{ 
   \label{some_def_845}
   \mathcal T^\mp_{i} = \int d^2z\,   {\cal V}^\mp_{i} \; .
 }  
Taking into account the freedom to fix three points on the world-sheet via the $SL(2,\mathbb C)$ symmetry, the three-tachyon scattering amplitude is  given by
\eq{
  \label{threetachyon}
  &\bigl\langle\, \mathcal T_1\: \mathcal T_2\:\mathcal  T_3\, \bigr\rangle^\mp 
  =  \int  \prod_{i=1}^3 d^2 z_i\: \delta^{(2)}(z_i-z_i^0)\, \vert
   z_{12}\,z_{13}\,z_{23}\vert^2  \, 
    \bigl\langle \,{\cal V}_1\, {\cal V}_2 \,{\cal V}_3\, \bigr\rangle^\mp \;,
}
where we have put the superscript $\mp$ indicating the $H$-flux and $R$-flux background
outside the bracket in order to shorten the notation.
Using the general formula \eqref{app_Ntach} given in  appendix
\ref{app_Ntachy}, we obtain for correlator of three vertex operators
\eq{
  \label{threetachyonb}
   \bigl\langle \,{\cal V}_1 \,{\cal V}_2 \,{\cal V}_3 \,\bigr\rangle^\mp
   =\frac{\delta(p_1+p_2+p_3)}{\vert z_{12}\,z_{13}\,z_{23}\vert^2}
       \exp\Bigl[ -i\hspace{0.5pt}\theta^{abc}\, p_{1,a} p_{2,b} p_{3,c}  \bigl[
  {\cal L}   \bigl( {\textstyle \frac{z_{12}}{z_{13}} }\bigr) \mp {\cal L}
   \bigl({\textstyle \frac{\ov z_{12}}{ \ov z_{13}}}\bigr)\bigr] \Bigr]_{\theta} \,,
}
where $[\ldots]_{\theta}$ indicates that the result is valid only up to linear order in $\theta$.
The full scattering amplitude then becomes
\eq{
  \label{threetacyon}
  &\bigl\langle\, \mathcal T_1\: \mathcal T_2\:\mathcal  T_3\, \bigr\rangle^\mp
  = \int \prod_{i=1}^3  d^2 z_i\, \delta^{(2)}(z_i-z_i^0) \, \delta(p_1+p_2+p_3) \times \\
 & \hspace{140pt}
 \exp\Bigl[ -i\hspace{0.5pt}\theta^{abc}\, p_{1,a} p_{2,b} p_{3,c}  \bigl[
  {\cal L}   \bigl( {\textstyle \frac{z_{12}}{z_{13}} }\bigr) \mp {\cal L}
   \bigl({\textstyle \frac{\ov z_{12}}{ \ov z_{13}}}\bigr)\bigr] \Bigr]_{\theta}.
}

Let us now study the behavior of \eqref{threetachyonb}  under
permutations of the vertex operators ${\cal V}^\mp_i$. 
Before applying momentum conservation, 
the three-tachyon amplitude for a permutation $\sigma$ of
the vertex operators can be computed 
using the  relations \eqref{sumdilogid}. 
With $\epsilon=-1$ for the $H$-flux and $\epsilon=+1$ for the $R$-flux, 
one finds\footnote{Exploiting  the freedom of adding integration constants to
\eqref{tpf_01}, we could have chosen 
$-{3L(1)\over 2}+{\cal L}\bigl( {\textstyle \frac{z_{12}}{z_{13}} }\bigr)$ for the basic holomorphic 
three-point function. Then, 
the phase in \eqref{phasethreeperm} would be vanishing also in the case of $R$-flux. However, such a constant term corresponds to a constant 
shift in  the Jacobi-identity \eqref{result}, implying it to 
be generically non-vanishing, which we consider to be unnatural. Therefore, comparing with the analogous  computation for the open string \cite{Chu:1998qz}, 
 we choose the integration constant to be zero  in the present situation.}
\begin{equation}
  \label{phasethreeperm}
  \bigl\langle \, {\cal V}_{\sigma(1)}   {\cal V}_{\sigma(2)}  {\cal V}_{\sigma(3)}  \bigr\rangle^\epsilon=
  \exp\Bigl[ \,i\left({\textstyle \frac{1+\epsilon}{ 2}}\right)  \eta_\sigma\,  \pi^2\,  \theta^{abc}\, p_{1,a} 
  \,p_{2,b} \,p_{3,c} \Bigr]
  \bigl\langle {\cal V}_1\,  {\cal V}_2\,  {\cal V}_3  \bigr\rangle^\epsilon \;,
\end{equation}
where in addition $\eta_{\sigma}=1$ for an odd permutation and 
$\eta_{\sigma}=0$ for an even one. 
Thus,  for the $R$-flux background a non-trivial phase may appear
which, in this paper, we have established up to linear order in the flux. 
This situation is similar to the open string where an analogous phase 
hinted towards a noncommutative star-product (see section \ref{sec_review_SW}).
Indeed, as will be discussed in more detail in section \ref{secthree}, the phase in \eqref{phasethreeperm} can be recovered from 
a three-product on the space of  functions
$V_{p_n}(x)=\exp( i\, p_n \cdot x )$, which can be defined as 
\eq{
  \label{threebracketexp}
   V_{p_1}(x)\,\tri\, V_{p_2}(x)\, \tri\, V_{p_3}(x)\stackrel{\rm def}{=} 
    \exp\Bigl( -i \,{\textstyle \frac{\pi^2}{2}}\, \theta^{abc}\,
   p_{1,a}\, p_{2,b}\, p_{3,c} \Bigr) V_{p_1+p_2+p_3}(x)\; .
}
However, in correlation functions operators
are understood to be radially ordered and so changing the order
of operators should not change the form of the amplitude.
This is known as crossing symmetry which is one of the defining
properties of a CFT and thus should also be satisfied
for our CFT$_H$. In the case of the $R$-flux background, this
is reconciled by applying momentum conservation leading to
\eq{
  p_{1,a} \,p_{2,b}\, p_{3,c}\,\theta^{abc} = 0
  \hspace{40pt}{\rm for}\hspace{40pt} p_3=-p_1-p_2 \;.
}
Therefore, scattering amplitudes of three tachyons do not receive
any corrections at linear order in $\theta$  both for the $H$- and
 $R$-flux.  We therefore find
\eq{
  \bigl\langle\, \mathcal T_1\: \mathcal T_2\:\mathcal  T_3\,
  \bigr\rangle^\mp=\delta(p_1+p_2+p_3) \; .
}
This is analogous to the situation in noncommutative open string theory, 
where the 
two-point function \eqref{correla} does not receive any corrections.


\subsection[$N$-tachyon amplitudes]{N-tachyon amplitudes}

In analogy to the result for the open string shown in equation \eqref{correla}, 
we now want to detect phases possibly appearing for the product of $N$ closed string
tachyon vertex operators. Before we consider the general case,
let us start with the amplitude of four tachyons.
Employing the general formula \eqref{app_Ntach}, up to linear
order in $\theta$ we obtain
\eq{
\label{fourptampl}
  &\bigl\langle {\cal V}_1\,{\cal V}_2\,{\cal V}_3\,{\cal V}_4 \bigr\rangle^\mp = 
  \bigl\langle V_1\,V_2\,V_3\,V_4 \bigr\rangle^\mp_0 \;\times \\
  &\hspace{100pt} \exp \biggl[  -i \hspace{0.5pt}
   \theta^{abc}\!\! \sum_{1\leq i<j<k\leq 4}  p_{i ,a} \,p_{j,b}\, p_{k,c}  
   \Bigl[ {\cal L}\bigl({\textstyle \frac{z_{ij}}{ z_{ik}}}\bigr)
  \mp  {\cal L}\bigl({\textstyle\frac{\ov z_{ij}}{ \ov z_{ik}}}\bigr)
  \Bigr] \biggr]_{\theta} \;.
}
Again, the difference between $H$- and $R$-flux is
given by the sign between the holomorphic and  the anti-holomorphic contribution, and 
the four-point function $\bigl\langle V_1\,V_2\,V_3\,V_4 \bigr\rangle^\mp_0$
is just the one from the free theory.
We can now  determine the behavior of the amplitude under a permutation 
of  the vertex operators. Prior to using momentum conservation, invoking
the fundamental identity of the Rogers dilogarithm \eqref{sumdilogid},  we 
again find momentum dependent phase factors. 
Analogous to the three-tachyon amplitude, these arise in the case of $R$-flux 
and  can be described as resulting from a deformed
four-product of the form
\eq{
  \label{fourbracketexp}
   &V_{p_1}(x)\,\tri_4\, V_{p_2}(x)\, \tri_4\, V_{p_3}(x) \, \tri_4\, V_{p_4}(x)\stackrel{\rm def}{=} 
   \exp\Bigl[-i\, {\textstyle \frac{\pi^2}{2}} \, \theta^{abc}\, \bigl(
   p_{1,a}\, p_{2,b}\, p_{3,c} \\
   &\hspace{100pt}
   +p_{1,a}\, p_{2,b}\, p_{4,c} + 
   p_{1,a}\, p_{3,b}\, p_{4,c} +
   p_{2,a}\, p_{3,b}\, p_{4,c}
   \bigr) \Bigr]\, V_{\sum p_i}(x)\; .
}
However, employing momentum conservation, one can show that this phase
becomes trivial so that the four-tachyon amplitude is indeed
 crossing symmetric.

This computation for four tachyons can straightforwardly be generalized to higher $N$-tachyon
amplitudes for which we find
\eq{
  &\bigl\langle {\cal V}_1\,{\cal V}_2\,\dots \,{\cal V}_N \bigr\rangle^\mp = 
  \bigl\langle V_1\,V_2\,\ldots\, V_N \bigr\rangle^\mp_0 \;\times \\
  &\hspace{93pt} \exp \biggl[  -i  \hspace{0.5pt}
   \theta^{abc}\!\! \sum_{1\leq i<j<k\leq N}  p_{i ,a} \,p_{j,b}\, p_{k,c}  
   \Bigl[ {\cal L}\bigl({\textstyle \frac{z_{ij}}{ z_{ik}}}\bigr)
  \mp  {\cal L}\bigl({\textstyle\frac{\ov z_{ij}}{ \ov z_{ik}}}\bigr)
  \Bigr] \biggr]_{\theta} \;.
}
The phase factors appearing when permuting two vertex operators for the case of the $R$-flux
background can be encoded in a deformed $N$-product  of the form
\eq{
\label{Nbracketexp}
   V_{p_1}(x)\,\tri_N\, \ldots\, \tri_N\, V_{p_N}(x) \stackrel{\rm def}{=} 
   \exp\Bigl(-i \,{\textstyle \frac{\pi^2}{2}}\, \theta^{abc}\!\!\!
   \sum_{1\le i < j <  k\le N} \!\!
   p_{i,a}\, p_{j,b}\, p_{k,c}   \Bigr)\; V_{\sum p_i}(x)\; .
}
The phase becomes again trivial after employing momentum conversation so that
all $N$-tachyon correlators are crossing symmetric. 
This signals that the basic principle of perturbative closed string theory,
namely conformal field theory, seems to be compatible with  
non-geometric backgrounds for which the $N$-product of functions
is deformed by \eqref{Nbracketexp}.


\subsection{The fluxed Virasoro-Shapiro  amplitude}
\label{sec_fluxedVS}

The four-tachyon scattering amplitude has played an 
important role in the history of string theory. In fact,
many aspects of the theory were detected by analyzing
its properties. 
In a similar spirit, we now further elaborate on the four-tachyon
scattering amplitude \eqref{fourptampl}.
Using  momentum conservation $\sum_i p_i=0$ to eliminate $p_4$, one obtains
\begin{equation}
\label{fourptc}
\begin{split}
  &\bigl\langle {\cal V}_1\, {\cal V}_2\, {\cal V}_3\, {\cal V}_4 \bigr\rangle^{\mp}  =
  \bigl\langle {V}_1\, {V}_2\, {V}_3\, {V}_4 \bigr\rangle^\mp_{0}\times \\
 & \hspace{25pt}
 \exp\biggl[-i \theta^{abc}\,   p_{1,a}\, p_{2,b}\, p_{3,c}\,  
  \Bigl[  {\cal L}({\textstyle \frac{z_{12}}{z_{13}}}) -  
 {\cal L}({\textstyle \frac{z_{12}}{ z_{14}}})+ 
 {\cal L}({\textstyle \frac{z_{13}}{ z_{14}}}) -
 {\cal L}({\textstyle \frac{z_{23}}{ z_{24}}})
 \mp {\rm c.c.} \Bigr]\biggl]_{\theta} .
  \end{split}
\end{equation}
Next, we simplify the sum over the four ${\cal L}$-functions
by using the five-term relation of the (complex) Rogers dilogarithm. 
Unfortunately, as indicated in appendix \ref{dilog}, this 
relation becomes quite involved in the complex case
since  logarithmic corrections of the form $F(z_{ij})=\sum \log z_{ij}$ appear.
However, observing that they satisfy 
$\partial_i \partial_j \partial_k F(z_{mn})=0$
for all $i,j,k\in \{1,2,3,4\}$ and $i\ne j\ne k\ne i$,
they  appear not to be physical.
These corrections  are closely related to the existence of branch cuts and are very 
effectively described by the so-called {\it extended  Rogers dilogarithm}.
As shown in detail in appendix \ref{dilog}, this allows us to express
the four-point function \eqref{fourptc} in terms of
${\cal L}(X)$, where the cross-ratio $X$ is defined as
\begin{equation}
 X=\frac{(z_1-z_2)(z_3-z_4)}{(z_1-z_3)(z_2-z_4)}\; .
\end{equation}
As it is also shown in appendix \ref{dilog},
requiring  that the four-tachyon correlator is crossing symmetric
leads  to 
the $SL(2,\mathbb C)$ invariant and explicitly crossing symmetric 
 four-tachyon amplitude 
\eq{
\label{fourtachyons}
 &\langle {\cal T}_1\, {\cal T}_2\, {\cal T}_3\, {\cal T}_4\rangle^{\mp}=\\
 &\hspace{40pt}
 \int d^2 X\:  {\exp\Bigl[-i\hspace{0.5pt} \theta^{abc}\,   p^1_a\, p^2_b\, p^3_c\,  
     \bigl[ (-{3\over 2}L(1) + {\cal L}(X)) \mp (-{3\over 2}L(1) + {\cal L}(\ov X)) \bigr] \Bigr]_{\theta} \over
     \vert X\vert^{2-2a}\; \vert 1-X\vert^{2-2c} } \;,
}
where the integrated tachyon vertex operator $\mathcal T_i^{\mp}$ was defined in \eqref{some_def_845} and where we employed
\eq{
  \label{some_def_6329}
     a={\alpha'\over 4}(p_1+p_4)^2-1 \;, \hspace{16pt}
     b={\alpha'\over 4}(p_1+p_3)^2-1 \;, \hspace{16pt}
     c={\alpha'\over 4}(p_1+p_2)^2-1\;.
}
The three corresponding Mandelstam variables read $u=-(p_1+p_4)^2$,
$t=-(p_1+p_3)^2$ and $s=-(p_1+p_2)^2$, and the on-shell external tachyons
satisfy $\alpha' p_i^2=4$ so that $a+b+c=1$. 
We consider \eqref{fourtachyons} as  the generalization of the Virasoro-Shapiro amplitude, 
which in our case includes corrections up to first order in $\theta$.
We therefore call it the (linearized) fluxed Virasoro-Shapiro (FVS) amplitude.

The usual Virasoro-Shapiro amplitude is just its $\theta=0$ limit,
in which case the integral can be solved in closed form. The result reads 
\eq{
  \bigl\langle T_1\, T_2\, T_3\, T_4 \bigr\rangle_{0}=
    2\pi\, {\Gamma(a)\, \Gamma(b)\, \Gamma(c)\over \Gamma(a+b)\,
        \Gamma(a+c)\, \Gamma(b+c)}\; ,
}
which, as it is well-known, has single poles at $a,b,c=-n$ for $n\in \mathbb Z^+_0$ 
corresponding to  the Regge excitations of mass $m_n^2=\frac{4}{\alpha'} (n-1)$.
Furthermore, unitarity implies that on a resonance $R_n$ of mass $m$, the four-tachyon
amplitude factorizes into two three-point functions
\eq{
  & \langle T_1\, T_2\, T_3\, T_4\rangle_{0} 
    \overset{\ u\to\,  -m^2}\simeq \\
  &\hspace{50pt}    { \langle\, T(p_1)\, T(p_4)\, R_n(-p_1-p_4)\,\rangle_{0}\, 
   \langle\,  T(p_2)\, T(p_3)\, R_n(-p_2-p_3)\, \rangle_{0} \over
       (p_1+p_4)^2+m_n^2 } \;.
}   
This factorization does not only occur in the $u$-channel, as presented, but similarly also in the $s$- and $t$-channel.

In the same way as the Virasoro-Shapiro amplitude contains
information about the underlying theory, we expect that
also  \eqref{fourtachyons} does
contain  new information on the  scattering
of four tachyons in a small three-form flux background.  Ideally,
one would solve the integral in a closed form, for instance
by the method of Kawai-Lewellen-Tye (KLT) \cite{Kawai:1985xq}.
However, even without having such an  explicit result at our
disposal, we can proceed and analyze the pole structure of the amplitude.
Thus, let us consider the correction linear in the flux which reads
\eq{
  \label{fourtachyonslin} 
  &\delta_1\bigl\langle {\cal T}_1\, {\cal T}_2\, {\cal T}_3\, {\cal T}_4\bigr\rangle^{\mp}  \\
  &\hspace{20pt}
  =-i\hspace{0.5pt} \theta^{abc}\,   p_{1,a}\, p_{2,b}\, p_{3,c}
 \int d^2 X\,  {  
      \bigl[-{3\over 2}L(1) + {\cal L}(X)\bigr] \mp \bigl[-{3\over 2}L(1) + {\cal L}(\ov X)\bigr]  \over
     \vert X\vert^{2-2a}\; \vert 1-X\vert^{2-2c} }\, .
}
We are interested in corrections to  the (former) resonances
at $u={4\over \alpha'} (n-1)$. These divergences originate from the
region $|X|<1$  and can be computed
by first introducing polar coordinates $(r,\varphi)$ and then
expanding the integrand as a series in $r$ around the origin.
In this respect it is useful to invoke the relation (see the end of appendix \ref{dilog})
\begin{equation}
\label{fourptf}
  {\cal L}(x)=3\hspace{0.5pt} L(x)\mp 
    {i\pi\over 2} \log\left(x(1-x)\right)\;, 
    \hspace{50pt}
    x\in\mathbb C \;,
\end{equation}
with the minus sign corresponding to the upper half plane, ${\rm Im}(x)>0$,  and the
plus sign to the lower half plane ${\rm Im}(x)<0$.
Moreover, besides the basic relation $\; \log X= \log r + i\varphi\,$, 
it is easy to verify the following  expansions
\eq{
   \begin{array}{ll}
   \displaystyle {\rm Re}\bigl(\log(1-X)\bigr)=-\sum_{n=1}^\infty {r^n\over n} \cos(n\varphi) \;, 
   \hspace{12pt}
   &
   \displaystyle {\rm Im}\bigl(\log(1-X)\bigr)=-\sum_{n=1}^\infty {r^n\over n} \sin(n\varphi) \;,  
   \\
   \displaystyle {\rm Re}\bigl({\rm Li}_2(X)\bigr)=\sum_{n=1}^\infty {r^n\over n^2} \cos(n\varphi) \;,
   &
   \displaystyle  {\rm Im}\bigl({\rm Li}_2(X)\bigr)=\sum_{n=1}^\infty {r^n\over n^2} \sin(n\varphi)\;.
    \end{array} 
    \\[-3mm]
}
It remains to expand the $1/|1-X|^{2-2c}$ term, for which we find
\eq{   
  {1\over |1-X|^{2-2c}}=&\sum_{n=0}^\infty a_n(\varphi)\, r^n \\
     =&\:1+\bigl[-2(c-1)\cos(\varphi)\bigr]\, r 
       +\bigl[ (c-1)+2(c-1)(c-2)\cos^2(\varphi)\bigr]\, r^2 \\
       &+ \bigl[ -2(c-1)(c-2)\cos(\varphi) - {\textstyle {4\over 3}} (c-1)(c-2)(c-3)
          \cos^3(\varphi)\bigr]\, r^3 \\
          &+ O(r^4)\; . 
          \\[-3mm]
}


\subsubsection*{Pole structure for $H$-flux}
\label{sec_poleforh}

With the help of the above expressions, we can now analyze the pole structure of the FVS-amplitude
at linear order in $\theta^{abc}$. We start with the case of  $H$-flux which corresponds to the minus sign in \eqref{fourtachyonslin}.
The $r$-expansion at linear order
in $\theta$ reads
\eq{
\label{fourtachyonstotalima}
  \hspace{125pt}&\hspace{-125pt}
  \delta_1\bigl\langle {\cal T}_1\, {\cal T}_2\, {\cal T}_3\, {\cal T}_4\bigr\rangle^{-}  
  =2\hspace{0.5pt} \theta^{abc}\,   p_{1,a}\, p_{2,b}\, p_{3,c}
 \int_0^1\! dr\! \int_{-\pi}^{+\pi} \!d\varphi\; r^{2a-1} \sum_{n=0}^\infty a_n(\varphi)\, r^n\:\times \\
 \biggl[ \hspace{25pt}
   3&\sum_{n=1}^\infty {r^n\over n^2} \,\sin(n\varphi)-{3\over 2} \log (r) 
    \sum_{n=1}^\infty {r^n\over n} \sin(n\varphi) \\
   -{3\over 2} \,\varphi
  &\sum_{n=1}^\infty {r^n\over n} \cos(n\varphi) \mp {\pi\over 2} \log (r)
  \pm{\pi\over 2} \sum_{n=1}^\infty {r^n\over n} \cos(n\varphi)\biggr] ,
}
where again the upper sign holds in the upper
half-plane $0\le \varphi\le \pi$ while the lower sign indicates the
lower half-plane $-\pi\le \varphi\le 0$.
We have furthermore restricted the integration to a disk of unit radius around
the origin in the complex plane. This allows us to read off the 
poles in $a={\alpha'\over 4} (p_1+p_4)^2-1$ since $\int_0^1 dr\, r^{2a+n-1}=(2a+n)^{-1}$ for
$n\ge 0$.

At each order in $r$, the above integral trivially vanishes since the
integrand is anti-symmetric in $\varphi$. Thus, at linear order in $H$
there are no corrections to the exchange modes at a particular  resonance $R_n$. 
For the tachyon  exchange,
this is consistent with what we found from the
three-point amplitudes in section \ref{sec_threetac}.
We will discuss the origin for the absence of corrections 
at the end of this section.


\subsubsection*{Pole structure for $R$-flux}

We now perform the same computation for the case of non-geometric $R$-flux, in which case the lower sign in \eqref{fourtachyonslin} applies.
We then find the following expansion
\eq{
\label{fourtachyonstotalc}
  \hspace{65pt}&\hspace{-65pt}
  \delta_1\bigl\langle {\cal T}_1\, {\cal T}_2\, {\cal T}_3\, {\cal T}_4\bigr\rangle^{+}  
  =-2i\hspace{0.5pt} \theta^{abc}\,   p_{1,a}\, p_{2,b}\, p_{3,c}
 \int_0^1\! dr\! \int_{-\pi}^{+\pi} \!d\varphi\; r^{2a-1} \sum_{n=0}^\infty a_n(\varphi)\, r^n\:\times \\
 \biggl[ \; &- \frac{3}{2}\, L(1)+
   3\sum_{n=1}^\infty {r^n\over n^2} \,\cos(n\varphi)
   -{3\over 2} \log (r) 
    \sum_{n=1}^\infty {r^n\over n} \cos(n\varphi) \\
   &+{3\over 2} \,\varphi
  \sum_{n=1}^\infty {r^n\over n} \sin(n\varphi) 
  \pm {\pi\over 2} \,\varphi
  \mp{\pi\over 2} \sum_{n=1}^\infty {r^n\over n} \sin(n\varphi)\hspace{40pt}\biggr] .
}
In this situation, at linear order in the flux there are non-vanishing contributions.
Let us discuss the first three leading terms in some more detail:

\pagebreak


\vspace{0.5cm}
\noindent
\underline{a) The tachyon pole}
\vspace{2mm}

\noindent
The first potential pole appears at order $r^{2a-1+n}$ for $n=0$, which after integration gives $(2a)^{-1}$ and thus corresponds to tachyon $(p_1+p_4)^2={4\over \alpha'}$. 
The $\varphi$ integral in this case is evaluated as
\eq{ 
  \int_{-\pi}^\pi \!d\varphi \left(-{3\over 2} L(1) \pm {\pi\over 2}
  \varphi\right)&= 2\int_0^\pi \!d\varphi \left( -{\pi^2\over 4}+
  {\pi\over 2}  \varphi\right)=0\; .
}
Therefore, by factorization the three-tachyon amplitude must vanish at linear
order in $\theta$, which is consistent with the result 
obtained by direct computation in the previous  section, and with
the implicit assumption to put the external tachyons
on the mass-shell $m^2=-{4\over \alpha'}$.


\vspace{0.5cm}
\noindent
\underline{b) A new tachyonic  pole}
\vspace{2mm}

\noindent
At order $r^{2a-1+n}$ with $n=1$ the integration gives $(2a+1)^{-1}$ corresponding to 
a resonance which, due to level matching,  was not part of the physical 
spectrum of the $26$-dimensional free bosonic string.
After some steps of computation we obtain
\eq{
\label{newtacmode}
\delta_1\bigl\langle {\cal T}_1\, {\cal T}_2\, {\cal T}_3\, {\cal T}_4\bigr\rangle^{+}  
    &\overset{\ u\to\,  -{2\over \alpha'}}\simeq
   -2i\hspace{0.5pt} \theta^{abc}\,   p_{1,a}\, p_{2,b}\, p_{3,c}\:\pi\:  {b-c  \over 2a+1}\\
   &\quad \simeq \quad\hspace{1.75pt}
   -2i\hspace{0.5pt} \theta^{abc}\,   p_{1,a}\, p_{2,b}\, p_{3,c}\:\pi\: {p_{14}\cdot p_{23}
        \over (p_1+p_4)^2-{2\over \alpha'}}\; ,
}   
where we note that $a,b,c$ were defined in \eqref{some_def_6329} and
where we employ $p_{ij}=p_i-p_j$.
As a qualitatively new feature, the mass
shift is {\it not continuous} because, once we turn on $\theta$ even only infinitesimally,
the new mass level appears. Via level matching, this mode appears to be a 
$\mathbb Z_2$ twisted mass state. 
It would be very interesting to understand the origin of this
mode in more detail. 
We may speculate that a former graviton mode
becomes tachyonic or that unphysical modes  becomes
physical due to the flux.
In any case, this new tachyon  signals  an instability of
the system, which we will discuss in more detail in section \ref{sec_tachy}.
In the following, we call this new tachyon a tachyon of type I. 


\vspace{0.5cm}
\noindent
\underline{c) The graviton pole}
\vspace{2mm}

\noindent
At order $r^{2a-1+n}$ for $n=2$ the exchange particle is the graviton. One obtains various
contributions, but the qualitatively new feature is an integral
involving a $\log( r)$ term. Anticipating the result for higher orders in $\theta$, one 
is led to integrals involving $(\log r)^m$ with $m>1$.\footnote{Here we are assuming that the corrections to tachyon correlators at higher order in the flux indeed give the exponential shown in \eqref{fourtachyons} beyond linear order.}
  Integrals
of this type can be computed explicitly as follows
\eq{
      \int_0^1 dr \,  r^{2a+n-1} (\log r)^m= {(-1)^m\,  m!\over (2a+n)^{m+1}}\; ,
}
and lead to higher order poles in the Mandelstam variables.
This seems to be in contradiction with the interpretation in terms
of Regge resonances. However,  a mass renormalization of such 
a Regge resonance would  indeed induce higher order poles of the form
\eq{
         {1\over p^2+m^2+(\Delta m)^2}={1\over p^2+m^2}-{(\Delta m)^2\over 
            (p^2+m^2)^2} +\ldots\; .
}
The total contribution from  the graviton exchange 
can then be expressed as 
\eq{
\label{gravipole}
\delta_1\bigl\langle {\cal T}_1\, {\cal T}_2\, {\cal T}_3\, {\cal T}_4\bigr\rangle^{+}  
    &\overset{\ u\to\,0}\simeq
   i\hspace{0.5pt} \theta^{abc}\,   p_{1,a}\, p_{2,b}\, p_{3,c} 
     \left( {9\pi(b-c)  \over 8(a+1)} +{3\pi(b-c)\over 8(a+1)^2}\right)\\
   &\ \; \simeq  \hspace{6pt}
   i\hspace{0.5pt} \theta^{abc}\,   p_{1,a}\, p_{2,b}\, p_{3,c} \left( {\pi\, p_{14}\cdot p_{23}
       \over (p_1+p_4)^2} + {3\pi\, p_{14}\cdot p_{23}
        \over (p_1+p_4)^4} \right)\; .
}   
We observe that the first simple pole  has the same momentum dependence
as the new tachyonic mode in \eqref{newtacmode}.
Due to the mass shift, also these modes
can become tachyonic, where in contrast to the tachyon
discussed in the last paragraph here the mass-shift 
{\it is continuous} in the parameter $\theta$.
Furthermore, we recall the computation of  $L_0$ eigenvalues \eqref{masssplit} for
the graviton vertex operators which are in accordance with the fact that
a small $R$-flux induces a light tachyonic mode. In the following
such tachyons are called of type II.


\vspace{0.5cm}
\noindent
\underline{d) Higher order poles}
\vspace{2mm}

\noindent
Finally,  the pole structure for higher Regge excitations is very similar.
At each former pole at $a=-n$ one now also finds a double pole and
there appears  a new pole at $a=-n+{1\over 2}$.
The general structure  can be expressed as
\eq{
 &\delta_1\bigl\langle {\cal T}_1\, {\cal T}_2\, {\cal T}_3\, {\cal T}_4\bigr\rangle^{+}  
    \simeq
   i\hspace{0.5pt} \theta^{abc}\,   p_{1,a}\, p_{2,b}\, p_{3,c}  \sum_{n=1}^\infty    
      \biggl( 
  {(b-c)\, P^{2n-2}_1(b,c)  \over (a+n-{1\over 2})} +   \\[0.2cm]
  &\hspace{150pt}{(b-c)\, P^{2n-2}_2(b,c)  \over (a+n)}+
  {(b-c)\, P^{2n-2}_3(b,c)  \over (a+n)^2}\biggr) \;,
}   
where $P^{2n-2}_i(b,c)$, $i=1,2,3$ are polynomials of order $2n-2$ in
$b$ and $c$. Thus, we see that the appearance of the aforementioned new $\mathbb Z_2$ 
twisted poles is generic.


\subsubsection*{Discussion of the pole structure}

Let us discuss the form of the poles obtained in b) and c) for
the $R$-flux case in more
detail, where part of it will  be at  a rather qualitative level
and might  turn out to be too naive.

First, we observe that the numerators for the single
and the double pole at linear order in $\theta$ 
have the same external momentum dependence.
This provides reason to believe that all these terms are related
to  massless gravitons $G$, dilaton $D$ and Kalb-Ramond fields $B$ at zeroth order in the flux. 
 Let us try to understand which
corrections to the two-tachyon--one graviton three point function 
$\langle {\cal T} {\cal T} {\cal G}\rangle$ can induce
these poles in the factorization limit.
At zero order in $\theta$, that is
for the free theory, it is known that  
$\langle {\cal T}_1\, {\cal T}_2\, {\cal G}_3\rangle^{\mp}\simeq
  g_{ab} \; p^a_{21}\, p^b_{21}$, i.e. there is
only a non-vanishing contribution for gravitons and for the dilaton.
Now let us assume that at linear order there is a correction of the form
\eq{
\label{scattTTGc}
 \langle {\cal T}(p_1)\, {\cal T}(p_2)\, {\cal G}(p_3)\rangle^{\pm}\simeq
  g_{ab} \; p^a_{21}\, p^b_{21}+
     p^a_{21}\, \theta^{b cd} p_{1,c} p_{2,d}\;
   \begin{cases} g_{ab}\quad & R{\rm -flux} \;,\\
                 b_{ab} & H{\rm -flux}\;, \end{cases}
}
and discuss its consequences for the four-tachyon amplitude.
\begin{itemize}

\item The two cases in \eqref{scattTTGc} imply that at linear order in $\theta$, 
for $H$-flux  the  $\langle TTB\rangle$ three-vertex is corrected 
and for $R$-flux the $\langle TTG\rangle$ and $\langle TTD\rangle$ 
vertices.
Thus, in the factorization limit of the four-tachyon amplitude,
up to linear order in $\theta$ we find
\eq{
\label{threegravi}
  \langle {\cal T}_1\, {\cal T}_2\, {\cal T}_3\, {\cal T}_4\rangle^{\mp} 
\overset{\ u\to\,0}\simeq  \begin{cases} {[p_{14}\cdot p_{23}]^2 \over \alpha' u}+ {(\theta^{abc}
  p_{1,a}p_{2,b}p_{3,c})\,  (p_{14}\cdot p_{23})\over \alpha' u} \;  & R{\rm -flux} \;,\\[0.2cm]
      {[p_{14}\cdot p_{23}]^2\over \alpha' u} & H{\rm -flux}\; , \end{cases}
}
where $u=-(p_1+p_4)^2$ is a Mandelstam variable.
In the case of non-geometric $R$-flux, we find  the 
single graviton pole shown in \eqref{gravipole}, while for $H$-flux we do 
not find any linear correction to the
graviton pole consistent with \eqref{sec_poleforh}.
It is  due this property that we included the $B$-field
in \eqref{scattTTGc} in the first place so that the expression  becomes
more symmetric. We will see in  a moment that, in view of T-duality,
it makes sense that the r\^ole of zero order graviton and
Kalb-Ramond field fluctuations  are exchanged 
by switching from   $R$-flux to  $H$-flux. 

\item As said, we expect the double pole in \eqref{gravipole} to arise
from a mass shift $\Delta m^2(g_{ab})$ of some of the
longitudinal gravitons/dilaton at linear order in $\theta$ which corresponds to the
 diagram shown in figure \ref{figmass}.
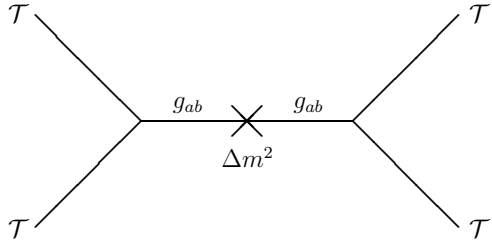
\begin{figure}[b]
\centering
\vskip5pt
\scalebox{0.8}{
\begin{picture}(200,100)
\thicklines
\put(0,0){\line(1,1){50}}
\put(0,100){\line(1,-1){50}}
\put(50,50){\line(1,0){100}}
\put(200,0){\line(-1,1){50}}
\put(200,100){\line(-1,-1){50}}
\put(93,43){\line(1,1){14}}
\put(107,43){\line(-1,1){14}}
\put(-13,-5){$\mathcal T$}
\put(-13,97){$ \mathcal T$}
\put(205,-5){$ \mathcal T$}
\put(205,97){$ \mathcal T$}
\put(65,56){$ g_{ab}$}
\put(122,56){$ g_{ab}$}
\put(88,28){$\Delta m^2$}
\end{picture}
}
\vskip5pt
\caption{\small{Mass correction of the graviton in the $\langle \mathcal T_1\mathcal T_2\mathcal T_3\mathcal T_4\rangle$ amplitude.}}
\label{figmass}
\end{figure}
Since the factorization limit of the four-tachyon amplitude
involves a sum over all polarizations of the graviton,
it is not obvious what  the separate eigenvalues 
$\Delta m^2(g_{ab})$ are.\footnote{A  more detailed analysis 
of the off-diagonal terms
in the logarithmic OPE \eqref{OPE_TG} might allow to determine
the precise eigenvalues.}
But, interpreting the $R$-flux  as a nonassociative deformation of ordinary
space, one could imagine that part of the former
general covariance symmetry is broken and the corresponding graviton mode
becomes massive.
For the case of $H$-flux there are no double poles but, as before, 
mass shifts of the longitudinal $B_{ab}$-fields can remain
undetected. In fact, due to breaking of conformal symmetry
at second order in $H$, we expect that also some
of the $B$-field modes become massive in the linear $H$-flux background.
Thus, as mentioned before, under T-duality the r\^ole of graviton
and $B$-field fluctuations seem to get exchanged.

\item Besides these mass shifts linear in $\theta$, the $R$-flux
amplitude has also shown a new tachyon of type I with half
the (negative) mass squared of the bosonic tachyonic ground state.
For the $H$-flux case, again
we do not directly see such a mode, but via T-duality  expect that
there also exists a similar tachyonic mode.

\end{itemize}
To summarize, from the factorization of the four-tachyon amplitudes
in the $R$-flux and $H$-flux background as well as from T-duality, 
we infer that in both
cases there are two new types of tachyonic modes, which we have denoted by
type I and type II.
For the case of $R$-flux,
these are expected to be related to  zero order  graviton/dilaton modes, 
whereas for $H$-flux we  expect them to be zero order  $B$-field modes.


\section{Asymmetric backgrounds and nonassociative geometry}
\label{sec_ab_nassgeom}

After having gained some insight on the structure
of scattering amplitudes, we now provide
a physical interpretation of some of the results.
First, we explain the appearance of the new tachyons of type I and type II 
in the spectrum and  discuss tachyon condensation. Our line of reasoning will
lead to the conjecture that non-geometric $R$-flux is related
to an asymmetric version of the CFT defined via the
representation theory of the Kac-Moody algebra  $\widehat{su}(2)_k$, 
i.e. the CFT of the WZW model.

Second, we give arguments for a new space-time structure responsible for
the appearance of constant phase factors
in the $N$-tachyon amplitudes (prior to invoking
momentum conservation).
More concretely, as already indicated in the previous section,
we  define a new tri-product inducing
a product of $N$ functions on a nonassociative space, which
makes the proposal that non-geometric $R$-flux 
is related to nonassociative geometry more precise.


\subsection{Speculations about  tachyon condensation}
\label{sec_tachy}

From our analysis of the 
the four-tachyon amplitude 
we concluded that there appear two types of
new tachyonic modes.
They are different in the sense that
for non-vanishing flux, 
the type I tachyons  received a discrete shift in the mass,
whereas for the  type II tachyons  the mass shift was continuous
in the three-form flux. 
The question now arises what kind of instabilities these
new tachyons indicate?


\subsubsection*{H-flux background}

Let us discuss the $H$-flux background
first. As mentioned in section \ref{sec_prereq}, a flat background
with a constant $H$-flux satisfies the 26-dimensional string equations
of motion only up to linear order in $H$. Thus, we expect that
fluctuations around this background will detect higher order corrections in $H$
and will therefore develop  tachyonic modes, i.e.  relevant operators
from the two-dimensional perspective,
that will induce a renormalization group flow of the theory 
towards  a truly conformal fixed point. 
One can immediately propose two candidate conformally invariant theories
to which the unstable theory might flow: 
\begin{itemize}

\item The theory should  certainly be able to flow back to the flat background
    with vanishing $H$-flux.
    
\item It is well known that for $H\ne 0$ there exists another conformally
    invariant theory, which is the $\widehat{su}(2)_k$ WZW model 
    corresponding to
    a constant $H$-flux through a three-sphere $S^3$.
    
\end{itemize}
It is now tempting to speculate  that the two kinds of tachyons mentioned above
correspond
to these two  conformal field theories.
In this case,  the type I tachyon should be identified with the  
$\widehat{su}(2)_k$ model
and the type II tachyon with the trivial background.


\subsubsection*{R-flux background}

Given the discussion for the $H$-flux background, 
we may ask about  the analogous structure for the
tachyons in the non-geometric $R$-flux background.
The type II tachyon is again expected to lead
to a flow towards the trivial theory with vanishing $R$-flux.
However, the  type I tachyon is expected to induce a flow
towards a theory which  is  isomorphic 
to the $\widehat{su}(2)_k$ CFT.
In view of the current algebra  \eqref{def_cur_04}, a natural
candidate may be provided by an {\em asymmetric $\widehat{su}(2)_k$  model}.
The only difference to the usual symmetric  $\widehat{su}(2)_k$ WZW model 
is a flip in the relative sign of the chiral and anti-chiral
symmetry algebras. More concretely, 
with the plus sign standing for the symmetric $\widehat{su}(2)_k^{++}$ model and the
minus sign denoting the asymmetric $\widehat{su}(2)_k^{+-}$ model,
the chiral and anti-chiral
 Kac-Moody algebras read
\eq{  
   \bigl[j^a_m,j^b_n\bigr]&=+ i f^{ab}_c\,  j^c_{m+n} + k\, m\, \delta^{ab}
    \, \delta_{m+n} \;,\\[0.1cm]
    \bigl[\ov j^a_m,\ov j^b_n\bigr]&=\pm i f^{ab}_c\,  \ov j^c_{m+n} + k\, m\, \delta^{ab}
   \,  \delta_{m+n} \;.
}

When constructing highest weights representations for these two theories,
the only difference is that the anti-holomorphic raising and lowering operators
are defined differently. In particular, we find
\eq{ 
   &\widehat{su}(2)_k^{++}\, :\ \bigl\{J^3_m,J^\pm_m=J^1_m\pm i J^2_m\bigr\}\times
      \bigl\{\ov J\vphantom{J}^3_m,\ov J\vphantom{J}^\pm_m
      =\ov J\vphantom{J}^1_m\pm i \ov J\vphantom{J}^2_m \bigr\}\;, \\[1mm]
     &\widehat{su}(2)_k^{+-}\, :\  \bigl\{J^3_m,J^\pm_m=J^1_m\pm i J^2_m\bigr\}\times
      \bigl\{\ov J\vphantom{J}^3_m,\ov J\vphantom{J}^\pm_m
      =\ov J\vphantom{J}^2_m\pm i \ov J\vphantom{J}^1_m\bigr\}\; .
}
One therefore obtains the same representations, characters and modular
invariant partition functions for both theories, so that on the level of the CFT
the  $\widehat{su}(2)_k^{++}$ and $\widehat{su}(2)_k^{+-}$  model are 
indistinguishable.
However, even though at the level of the conformal field theory 
there is no difference between these
two models,  the target-space interpretation is different,
namely one is geometric and the other is non-geometric. 
Our conjecture is that the non-trivial conformal fixed point of
the model with $R$-flux is the asymmetric $\widehat{su}(2)_k^{+-}$ model. 
This would imply  an intricate relationship between
{\it $R$-flux and asymmetric string vacua}. We have to admit that
in our situation the asymmetry is barely visible, but in more
general cases it should be more apparent.


\subsection{A tri-product}
\label{secthree}

In the previous sections, we have computed scattering amplitudes of $N$ tachyons
and in the case of constant $R$-flux,  have detected relative phase
factors between different orders of insertion of the vertex
operators. As   required by conformal symmetry or
crossing symmetry, respectively, (at linear
order in $\theta$) these phases became trivial after invoking
momentum conservation.
In this section, we show that these relative phase factors
can be rephrased in terms of a generalization of the
Moyal-Weyl star-product, which we call a tri-product.

In particular, the phase appearing in the three-point correlator \eqref{phasethreeperm}
indicates that we can define a three-product of functions
$f(x)$ in the following way\footnote{As repeatedly emphasized, our methods 
are  only reliable up to linear order in the flux parameter $\theta$.}
\eq{
\label{threebracketcon}
   f_1(x)\,\tri\, f_2(x)\, \tri\, f_3(x) \stackrel{\rm def}{=} \exp\Bigl(
   {\textstyle {\pi^2\over 2}}\, \theta^{abc}\,
      \partial^{x_1}_{a}\,\partial^{x_2}_{b}\,\partial^{x_3}_{c} \Bigr)\, f_1(x_1)\, f_2(x_2)\,
   f_3(x_3)\Bigr|_{x} \;,
}
where we used the notation $(\ )\vert_x=(\ )\vert_{x_1=x_2=x_3=x}$.
Choosing $f_n(x)=\exp( i\hspace{0.5pt} p_n \cdot x )$ we obtain formula
\eqref{threebracketexp},
which after integration over $x$ gives
\eq{
\label{threebracketexpint}
   \int d^3x\;  f_1(x)\,\tri\, f_2(x)\, \tri\, f_3(x) &=  \exp\Bigl(-i {\textstyle {\pi^2\over 2}}\theta^{abc}\,
   p_{1,a}\, p_{2,b}\, p_{3,c} \Bigr)\,  \delta(p_1+p_2+p_3)\\
   &=\int d^3x\; f_1(x)\, f_2(x)\,  f_3(x)\; .
}
Note that \eqref{threebracketcon} is precisely the three-product anticipated in
\cite{Blumenhagen:2010hj}.
Indeed, the three-bracket for the coordinates $x^a$ can then be re-derived  as
the completely antisymmetrized sum of three-products
\eq{
\label{antisymtripcon}
   \bigl [x^a,x^b,x^c \bigr]=\sum_{\sigma\in P^3} {\rm sign}(\sigma) \;  
     x^{\sigma(a)}\, \tri\,  x^{\sigma(b)}\, \tri\,  x^{\sigma(c)} =
     3\pi^2\, \theta^{abc}\; ,
}
where $P^3$ denotes the permutation group of three elements.
In \cite{Blumenhagen:2010hj} this three-bracket was defined
as the Jacobi-identity of the coordinates, which can
only be non-zero if the space is noncommutative and nonassociative.

Next we consider the $N$-tachyon amplitude and the phase appearing
in equation \eqref{Nbracketexp}. This motivates us to define
the $N$-product
\eq{
   f_1(x)\, \tri_N\,  &f_2(x)\, \tri_N \ldots \tri_N\,  f_N(x) \stackrel{\rm def}{=} \\
   &\exp\left[ {\textstyle {\pi^2\over 2}} \theta^{abc}\!\!\!\!\! \sum_{1\le i< j < k\le N}
     \!\!\!\!  \, 
      \partial^{x_i}_{a}\,\partial^{x_j}_{b} \partial^{x_k}_{c} \right]\, 
   f_1(x_1)\, f_2(x_2)\ldots
   f_N(x_N)\Bigr|_{x} \;,
}
which is the closed string generalization of the open string
noncommutative product \eqref{Nbracketcon}.
This completely defines the new tri-product, which satisfies
the relation
\eq{
\label{thenicerel}
    f_1\,\tri_N\,  f_2  \,\tri_N\,  \ldots\,  \tri_N\, f_{N-1} \, \tri_N\, 1\,   \
   =  f_1\,\tri_{N-1}\,  \ldots\,  \tri_{N-1}\,  f_{N-1} \;.
} 
Specializing this expression  to $N=3$ gives
\begin{equation}
    f_1\,  \tri_2\,  f_2 = f_1\,\tri_3\, f_2\,\tri_3 \, 1= f_1 \cdot f_2  \; , 
\end{equation}
which just means that the tri-product of two functions
is the usual commutative point-wise product.
However, there are two main differences compared to the open 
string case.
\begin{itemize}

\item For the open string the star $N$-product was 
related to successive application of the usual Moyal-Weyl
bi-product.  
This simplifying behavior is not true for the tri-product, i.e.
the $N$-products $\ \tri_N$ {\it cannot} be related to successive
applications of the three-product $\ \tri\;\,=\;\,\tri_3$. 
For instance, for $N=5$ we find
\begin{equation}
   f_1\,\tri_5\, f_2\,\tri_5\, f_3 \,\tri_5\,  f_4\, \tri_5\,  f_5\ne
   \left[ f_1\, \tri\, f_2\, \tri\, f_3\right]\, \tri f_4\, \tri\,   f_5\; .
\end{equation}

\item In contrast to the open string case, the effect
of the tri-product in integrals vanishes, i.e. 
\eq{
\int d^nx\, f_1(x)\, \tri_N\,  f_2(x)\, \tri_N \ldots \tri_N\,  f_N(x)
=  \int d^nx\, f_1(x)\,  f_2(x)\,  \ldots   f_N(x)\; .
}
In other words, the difference between the tri-product and
the ordinary product  is a total derivative.
\end{itemize}
The first item means that one does not only have to specify a deformed product of three functions
(with the rest following), but has to specify a
definition for a deformed product of any number of functions. We suspect that
this a very general behavior of such NCA geometries.
The second item  means that closed strings can consistently
be defined on such nonassociative backgrounds, since  in string scattering amplitudes
it's effect vanishes, i.e. {\em closed on-shell strings are blind against this
deformation}.

\begin{figure}[ht]
\vskip10pt
\centering
\includegraphics[width=0.95\textwidth]{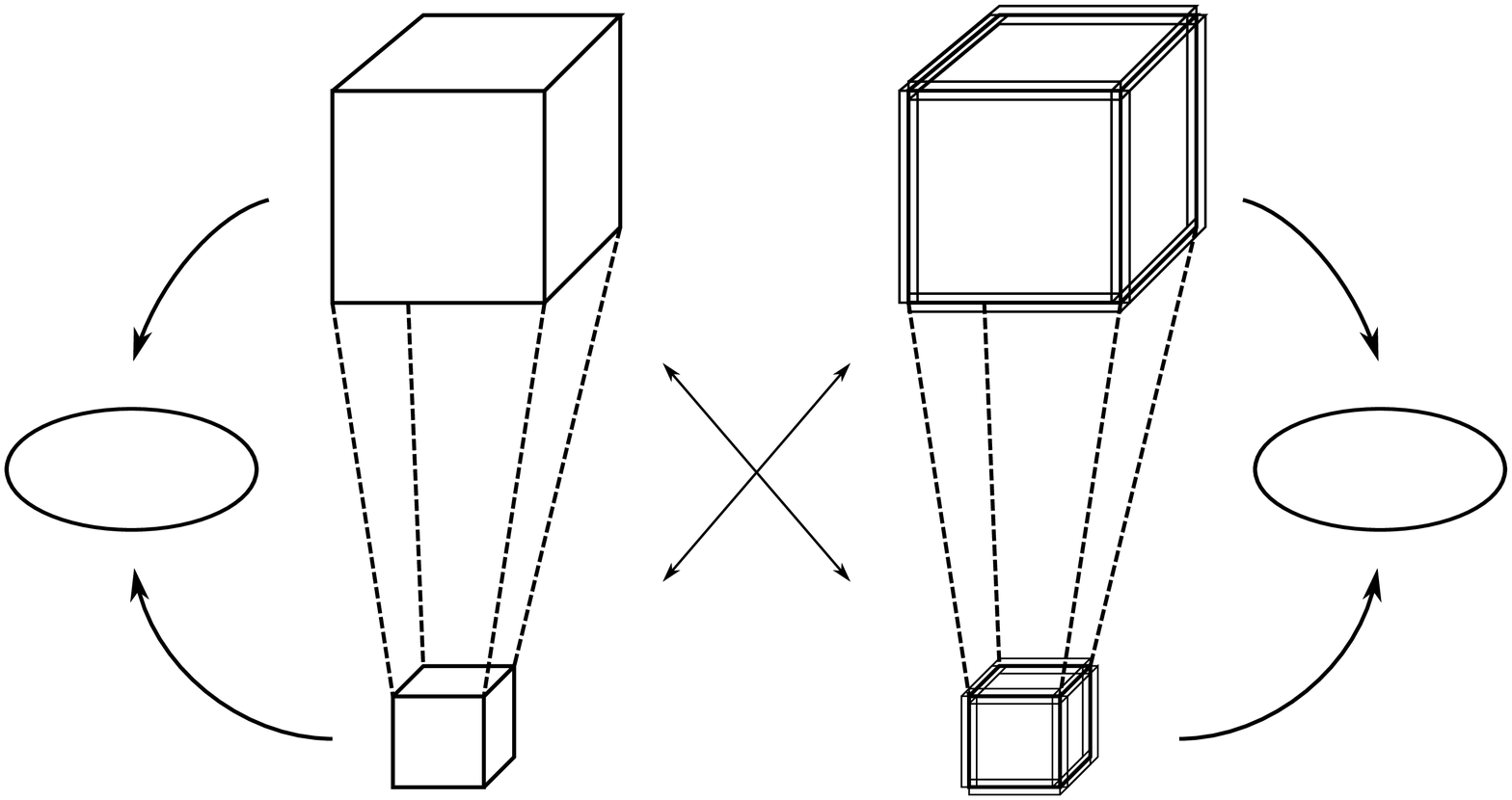}

\begin{picture}(0,0)
\put(-178,98){\footnotesize $\widehat{su}(2)^{++}_k$}
\put(148,98){\footnotesize $\widehat{su}(2)^{+-}_k$}
\put(-15,127){\footnotesize T-dual}
\put(-170,155){\footnotesize $\langle T_{\rm I}\rangle$}
\put(-170,40){\footnotesize $\langle T_{\rm I}\rangle$}
\put(157,155){\footnotesize $\langle T_{\rm I}\rangle$}
\put(157,40){\footnotesize $\langle T_{\rm I}\rangle$}
\put(-85,182){$\mathbb T^3$}
\put(-105,162){$\int H=k$}
\put(46,182){NCA $\mathbb T^3$}
\put(47,162){$\int \theta=k$}
\end{picture}
\caption{\small{T-duality relation between a $\mathbb T^3$-compactification with
constant flux $\int _{T^3} H=k$  and a nonassociative $\mathbb T^3$ with
constant $R$-flux $\int _{T^3} \theta=k$. Shown are also the proposed 
results of type I tachyon condensation, which do only depend on the quantized
flux and not the initial size of the $\mathbb T^3$.}}
\label{fig3}
\end{figure}

Let us finish our discussion  by mentioning that the mathematical analysis of 
such nonassociative spaces  is  beyond
the scope of this paper.\footnote{It would interesting to know
whether this tri-product is compatible with the rather abstract
notion of nonassociativity in \cite{Bouwknegt:2004ap}.}
We also would like to refer the reader to figure \ref{fig3}  where
the two  proposals
made in this section for  T-duality
and tachyon condensation of our initially 
 constant $H$-flux on flat space configuration are illustrated. 


\section{Conclusions}

In this paper, we have studied the structure of closed strings
moving in three-form flux backgrounds.
Our starting point was  a flat space-time with
 constant $H$-flux, which via T-dualities is related
to geometric and non-geometric flux backgrounds. 
In particular,  the $H$- and $R$-flux configurations 
were investigated up to linear order
in the flux in which they are expected to satisfy the 
string equations of motion.

Since our objection was to see whether a new nonassociative product
for the $R$-flux background can appear, we followed 
an approach analogously to the constant $B$-form background in the
open string case. 
We computed the three-point function of three coordinates $\mathcal X^a$ 
via conformal perturbation
theory and found that it can be expressed
in terms of the Rogers dilogarithm. This result is consistent
with what was derived as a limiting case of the
$SU(2)$ WZW-model in \cite{Blumenhagen:2010hj}.
Using this three-point function, we explicitly determined 
scattering amplitudes 
which revealed  a much more intricate structure  than
in the open string case. The reason is that for the open string
one can work with a free conformal field theory,
whereas for the closed string the theory is interacting and  only 
trustable (i.e. conformally invariant)   up to linear
order in the  fluxes. We explicitly derived
a number of OPEs and correlation functions of this CFT$_H$.

Furthermore, in the case of $R$-flux we found
relative phases factors in the $N$-tachyon amplitudes upon permutation of two
operators,
which vanish after applying momentum conservation.
We encoded the appearing  phases via a new nonassociative tri-product,
which generalizes the Moyal-Weyl product
to closed strings with $R$-flux and  supports
our proposal that nonassociative spaces 
are relevant for these backgrounds. 
This analysis showed that on-shell such a NCA deformation of
the target-space is compatible with the structure of
two-dimensional conformal field theory. However, off-shell 
the nonassociative structure should become
much more visible.

Moreover, we derived 
a conformally invariant and crossing symmetric
four-tachyon amplitude and studied its pole structure.
We observed the appearance of  two new types of
tachyons, for which we presented  an interpretation
in terms of the apparent instabilities of the system. 
In the case of $R$-flux we conjectured that the initial
model flows towards an asymmetric version of the WZW model.
Thus, non-geometric $R$-flux could be related
to left-right asymmetric string backgrounds, which from the
target space perspective seem to involve nonassociative geometries.
We believe that this points towards a coherent  picture for the appearance of
non-standard  geometric structures in string theory.
Also, recall from \cite{Blumenhagen:2000fp} that for the open string 
a non-trivial two-form flux can be
generated on a D-brane via an asymmetric rotation.
As just discussed, asymmetric solutions (like in the case of $R$-flux)   are 
similarly related to nonassociative bulk spaces probed by closed
strings.
From this perspective, commuting and associative geometries 
are rather the exception; generic solutions
of string theory should involve NC-brane  and NCA-bulk  geometries.

The analysis of the  CFT$_H$ theory at linear order in flux presented in this paper
clearly leads to a number of interesting issues worth to be studied. 
These include
\begin{itemize}

\item{A generalization from the bosonic string investigated in this paper to  the superstring, that is the construction of a SCFT$_H$.} 

\item{The OPEs for the graviton vertex operators show features 
reminiscent of logarithmic CFTs. It would be interesting 
to further analyze this structure.}
 
\item{Furthermore, the construction of the boundary CFT$_H$ is worth
  pursuing. In this case, the Freed-Witten anomaly \cite{Freed:1999vc} 
should be directly visible in the  boundary states since this anomaly is an effect  linear in the flux $H$.}

\end{itemize}

We close with a puzzling question. In our approach we treated
the non-geometric $R$-flux as a constant background. 
As we have illustrated, we expect that some of the
gravitons at zero order in the flux become  massive after taking into account
corrections at linear order. But, 
{\it to what kind of fluctuation do the antisymmetric polarizations
correspond to?} One possibility is that these are modes of the  $B$-field, though
they could also correspond to $R$-flux fluctuations implying
changes in the NCA parameter $\theta$.
In the latter case, one would 
promote the constant parameter $\theta^{abc}$ to a field $\Theta^{abc}(X)$
so that the nonassociative geometry is  defined by a non-vanishing
three-bracket of the form
\eq{
  \bigl[X^a,X^b,X^c \bigr] =  \Theta^{abc}(X)\; .
}
Conformal symmetry should then lead to on-shell equations of motions
for $\Theta^{abc}(X)$.


\vskip60pt

\subsubsection*{Acknowledgements}
We would like to thank Stefan F\"orste, Jaume Gomis, Igor Khavkine, Oliver
Schlotterer   and Andreas Wi\ss kirchen
for discussion
and Eran Palti for a very useful comment on $R$-flux.
R.B. thanks the University of Bonn for hospitality. D.L. likes to thank the theory department of CERN, where part of the work was done.
This research was also supported by the Munich Excellence Cluster for Fundamental Physics "Origin and the Structure of the Universe".
E.P. is supported by the Netherlands Organization for Scientific Research (NWO) under a FOM Foundation research program.

\vspace{1cm}


\clearpage
\appendix


\section{Appendix}

\subsection{T-dual flux backgrounds}
\label{Tdualbackgrounds}

In this appendix, we review some aspects of backgrounds with $H$-, geometric, non-geometric and $R$-flux.


\subsubsection*{H-flux background}

Following \cite{Kachru:2002sk,Shelton:2005cf,Wecht:2007wu},
let us start with a flat, rectangular three-torus  parametrized by coordinates $x^1$, $x^2$, $x^3$
with metric
\eq{
  ds^2 = R_1^2 \,\bigl(dx^1\bigr)^2 + R_2^2 \,\bigl(dx^2\bigr)^2 + R_3^2\,\bigl(dx^3\bigr)^2 \;.
}
The radii of $\mathbb T^3$ are denoted by $R_1,R_2,R_3$, and we  allow for a constant $H$-flux such that 
\begin{equation}\label{quant}
  \int_{T^3} H = N \in \mathbb{Z} \;.
\end{equation}
This implies  that the sigma-model equations of motion are  satisfied only at linear
order in $H$.  We can therefore consider this configuration as a consistent string background only up to this order. Furthermore, we are free to choose a gauge in which the $B$-field reads 
\eq{
  B_{x^2x^3} = N x^1\;.
}

As it turns out, it is useful to consider the above three-torus as a $\mathbb T^2$ in the $(x^2,x^3)$ direction fibered over an $S^1$ in the $x^1$ direction. This allows us to define a complex structure modulus $\tau$ and a K\"ahler modulus $\rho$ for $\mathbb T^2$ as
\eq{
  \tau = i\, R_2/R_3 \;, \hspace{50pt}  \rho(x^1) = N x^1 + i R_2 R_3\;,
}
which encodes \eqref{quant} as a parabolic monodromy 
\eq{
  \rho \to \rho + 2\pi R_1 N \hspace{40pt}{\rm when}\hspace{40pt} x^1 \to x^1 + 2\pi R_1\;.
}
The monodromy can be realized as a $SL(2,\mathbb{Z})$ transformation\footnote{To be more precise, we have to divide by an identification $2\pi R_1 \sim 1$. It is useful to think of each modulus as being an element of $\mathbb{C}P^1$ with $SL(2,\mathbb{Z})/\mathbb Z$ acting merely as matrix multiplication on the homogeneous coordinates.} on $\rho$, i.e. the moduli are preserved up to a $SL(2,\mathbb{Z})$ transformations when going once around the base circle.


\subsubsection*{Geometric flux background}

Let us now perform a T-duality transformation along one of the three isometric directions.\footnote{We should mention that \cite{Evans:1995su} describes a method for T-dualizing without referring to any Killing symmetries.} 
Utilizing the Buscher rules, we can perform a first T-duality in, say, the $x^3$ direction to obtain
\eq{
  \label{twto}
  ds^2 = R_1^2 \,\bigl(dx^1\bigr)^2 + R_2^2 \,\bigl(dx^2\bigr)^2 + \frac{1}{R_3^2} 
  \,\bigl( dx^3 + N x^1 dx^2 \bigr)^2	 
  \;,\hspace{40pt} B= 0 \;.
}
The complex structure and K\"ahler modulus are exchanged, that is $\tau' = \rho$ and $\rho' = \tau$, and we observe that the new metric receives an explicit dependence on the base-space coordinate $x^1$. In order for this metric to be globally well-defined we have to restore the periodicity along the base direction by identifying
\eq{
  \bigl(x^1,x^2,x^3\bigr) \sim \bigl(x^1 + 2\pi R_1\,,\,x^2\,,\,x^3 - 2\pi R_1 N x^2\bigr)\;,
}
which defines a so-called twisted torus.

In order to interpret the role of $N$ in this geometry, we introduce a dual basis of globally defined one-forms for $T(\mathbb T^3)$ as
\eq{
\eta^1 = dx^1 \;,\hspace{40pt}
\eta^2 = dx^2 \;,\hspace{40pt}
\eta^3 = dx^3 + Nx^1dx^2 \;. 
}
Employing Cartan's structure equation for a torsion-free connection  $\omega^a{}_b$
\eq{
\label{cse}
  d\eta^a = \eta^b \wedge \omega^a{}_b \; ,
}
for the above basis the only non-vanishing component of 
$\omega^a{}_b$ turns out to be
\eq{
  \omega^{x^3}{}_{x^1x^2} = -N \ ,
}
which is usually  referred to as geometric flux. However, in general Cartan's structure equation \eqref{cse} yields another constraint on $\omega^a{}_{bc}$ upon demanding $d^2\eta^a = 0$. The resulting  Jacobi identity $\omega^a{}_{b[c} \omega^b{}_{de]} = 0$ implies that  we can consider $\omega^a{}_{bc}$ to be structure constants of the Lie algebra.
Furthermore, for compact spaces one has to require in addition that 
$\omega^a{}_{ab} = 0$ (no sum) which is satisfied by  nilpotent algebras, and the resulting manifolds are called nilmanifolds.


\subsubsection*{Non-geometric flux background}

Since the metric \eqref{twto} does not  depend on $x^2$ explicitly, we are allowed to perform a second T-duality in the remaining direction of $\mathbb T^2$. Applying the Buscher rules once more yields
\eq{
  \label{Qmet}
  &ds^2 = R_1^2\,\bigl (dx^1\bigr)^2 + \frac{1}{R_2^2R_3^2 + N^2(x^1)^2}
  \,\bigl( R_3^2(dx^2)^2 + R_2^2(dx^3)^2 \bigr) \;,	\\
  &B_{x^2x^3} = -\frac{N x^1}{R_2^2R_3^2 + N^2(x^1)^2} \; .
}
The complex structure and K\"ahler modulus can be computed to be
\eq{
  \tau'' = i\,R_3/R_2 \;, \hspace{50pt}  \rho'' = -1/(Nx^1+iR_2R_3) \ ,
}
leading to the monodromy 
\eq{
  1/\rho'' \to 1/\rho''+2\pi R_1 N\hspace{40pt}{\rm when}\hspace{40pt}
  x^1 \to x^1 + 2\pi R_1\;.
}
This can still be realized as a parabolic $SL(2,\mathbb{Z})_{\rho}$ transformation. 
However, although the metric and the $B$-field are well-defined locally, it is not possible to describe them globally. The transition functions between local trivializations mix the $B$-field with the metric on the total space, which is known as  a $T$-fold \cite{Hull:2004in}. 
Note that in the present case, the parameter $N$ is related to the so-called non-geometric flux $Q_{x^1}{}^{x^2x^3} = N$.


\subsubsection*{R-flux background}

Finally, we can  consider a T-duality along the base direction $x^1$ which is however not captured by the Buscher rules as \eqref{Qmet} does not admit an isometry in this direction. But, in \cite{Shelton:2005cf,Wecht:2007wu} it is argued that although the background obtained by performing another T-duality seems to elude a geometric description even locally, it has to be included in a background independent formulation of string theory. 
We  characterize this background by a new type of flux, denoted by $R^{x^1x^2x^3} = N$.
It is obtained by formally applying the following chain of three T-duality transformations
\eq{
  H_{x^1x^2x^3} \;\xleftrightarrow{\;\; T_{x^3}\;\;}\;
   \omega_{x^1x^2}{}^{x^3} \;\xleftrightarrow{\;\; T_{x^2}\;\;}\;
  Q_{x^1}{}^{x^2x^3} \;\xleftrightarrow{\;\; T_{x^1}\;\;}\;
  R^{x^1x^2x^3} \; .
}


\subsection{The Rogers dilogarithm}
\label{dilog}

In this appendix, we summarize some properties of the complex Rogers dilogarithm and recall one of its generalizations defined in \cite{1997math.....12226N}. The latter will be used in the main text to rewrite the four-tachyon amplitude \eqref{fourptc} in a form which only depends on the cross-ratio 
\begin{equation}
\label{crossratio}
  z = \frac{(z_1 - z_4)(z_2-z_3)}{(z_1-z_3)(z_2-z_4)} \;,
\end{equation}
implying that this amplitude is manifestly invariant under $SL(2,\mathbb{C})$ transformations.
For a more detailed analysis of the mathematical aspects of the Rogers dilogarithm function we would like to refer the reader to  \cite{Zagier,Kirillov:1994en}, whereas its generalization is described in detail in \cite{1997math.....12226N, 2003math......7092N}.


\subsubsection*{Definition and fundamental properties}

The Rogers dilogarithm function $L(x)$ for real arguments $x$ is defined in the following way
\begin{equation}
\label{rogers} 
L(x) :=\textrm{Li}_2(x) + \frac{1}{2}\, \log(x)\log(1-x) \;,  \hspace{40pt} 0<x<1 \;,
\end{equation} 
where $\textrm{Li}_2(x)$ denotes the Euler dilogarithm function given by 
\begin{equation}
\label{euler}
\textrm{Li}_2(x) := \sum_{n=1}^{\infty} \:\frac{x^n}{n^2} = -\int_0 ^x \frac{\log(1-y)}{y} \;,
\hspace{40pt} 0\leq x \leq 1 \;.
\end{equation}
With the help of \eqref{euler}, the integral representation of the Rogers dilogarithm can be deduced as
\begin{equation}
\label{integraldarstellung}
L(x)= -\frac{1}{2} \int_0 ^x \left( \frac{\log(1-y)}{y} + \frac{\log(y)}{1-y}\right) dy \;.
\end{equation}
Furthermore, from these definitions one can derive two functional relations, which in turn uniquely characterize the Rogers dilogarithm function
\eq{
\label{app_Lrel_eral}
&L(x) + L(1-x) = L(1) \;, \\
& L(x) - L(y) + L\bigl({\textstyle \frac{y}{x}}\bigr) - L \bigl( \textstyle{\frac{1-x^{-1}}{1-y^{-1}}} \bigr) 
+ L\bigl( \textstyle{\frac{1-x}{1-y}} \bigr) = 0 \;.
}

Employing the integral representation \eqref{integraldarstellung}, one can analytically continue $L(x)$ to the domain $\mathbb{C} \setminus \{0,1\}$. However, the resulting function $L(z)$ is not single valued any more and one should use the universal cover of $\mathbb{C} \setminus \{0,1\}$ as the domain of definition.   
For the complex Rogers dilogarithm  the relation 
\eq{
\label{complfund}
  L(z)+L(1-z)=L(1) 
}
still holds, but the five-term relation in \eqref{app_Lrel_eral} receives logarithmic corrections.
To describe the systematics of those corrections, let us introduce the following generalization which is due to  Neumann \cite{1997math.....12226N}
\begin{equation}
\label{NeumannRogers}
R(z\,;p,q) := L(z) -\frac{\pi^2}{6} + \frac{\pi i }{2}\Bigl(p \log(z-1) + q \log
z \Bigr)\; .
\end{equation}
Here, $p,q$ are integer numbers and the constant is just a convenient normalization. For \eqref{NeumannRogers} it is then possible to establish again a five-term relation
\eq{
\label{5terma}
&R(x\,;p_0,q_0)-R(y\,;p_1,q_1)+R\bigl(\textstyle{\frac{y}{x}}\,;p_2,q_2\bigr) \\
& \hspace{100pt}
 -R\bigl(\textstyle{\frac{1-x^{-1}}{1-y^{-1}}}\,;p_3,q_3\bigr) + R\bigl(\textstyle{\frac{1-x}{1-y}}\,;p_4,q_4\bigr) = 0 \;,
}
where the integers $p_i,q_i$ have to obey some restrictions called \emph{flattening conditions}.


\subsubsection*{Five-term relation}

In the following, we sketch a geometric viewpoint on the arguments of the five-term relation (\ref{5terma}), but refer the reader to the original articles \cite{1997math.....12226N, 2003math......7092N}
for a more comprehensive treatment.

Let $z_1 \dots z_5$ be five distinct points in $\mathbb{C}\cup \{\infty\}$. If we choose four of them, we can interpret them as endpoints of a tetrahedron in hyperbolic three-space $\mathbb{H}^3$  located at $\partial \mathbb{H}^3 = \mathbb{CP}^1$, which is also known as an \emph{ideal} tetrahedron. Up to congruence, such an object is characterized by the cross-ratio of its four endpoints. For instance, omitting $z_5$ we have
\begin{equation}
   [z_1:z_2:z_3:z_4]:= \frac{z_{32}\, z_{41}}{z_{31}\, z_{42}}\; .
\end{equation}
Now we observe that, if we omit from $z_1 \dots z_5$ one point we get cross-ratio parameters corresponding to the arguments appearing in the five term relation \eqref{5terma}
\eq{
  \arraycolsep2pt
  \begin{array}{llllll}
   [z_2:z_3:z_4:z_5]  &=:& x \;,
   &[z_1:z_3:z_4:z_5]  &=:&  y \;,  \\[3mm]
   [z_1:z_2:z_4:z_5] &=& 
   \displaystyle \frac{y}{x}\;,
   &[z_1:z_2:z_3:z_5]  &=& 
   \displaystyle \frac{1-x^{-1}}{1-y^{-1}} \;, \\[3mm]
   [z_1:z_2:z_3:z_4] &=&
   \displaystyle  \frac{1-x}{1-y} \; .\hspace{40pt}
   \end{array}
}
This suggests a geometric reason for the five-term relation. Indeed, if we consider the so called scissors congruence group $\mathcal{P}(\mathbb{H}^3)$, which is given by the free $\mathbb{Z}$-module generated by three-dimensional polytopes in $\mathbb{H}^3$ modulo congruence relations (i.e. if we denote by $[P]$ the class of a polytope $P$, then we have $[P]= [P_1]+\dots+[P_n]$, if we get $P$ by gluing $P_1, \ldots, P_n$ along common faces), it turns out that in $\mathcal{P}(\mathbb{H}^3)$ we obtain the following relation
\begin{equation}
  [x]+\left[\frac{y}{x}\right]+\left[\frac{1-x}{1-y}\right] = [y]+\left[\frac{1-x^{-1}}{1-y^{-1}}\right] \;,
\end{equation}
where $[z]$ denotes the class of the tetrahedron with cross-ratio parameter $z$.
Now, as stated in proposition 4.5 of \cite{1997math.....12226N}, it turns out that a similar relation
holds also on the domain of definition of the extended version
\eqref{NeumannRogers} of the Rogers dilogarithm, if the  additional parameters
obey the so-called flattening condition. For instance, if  all the arguments  are in the
upper complex half-plane this condition is given by the following set of linear equations
\eq{
  \arraycolsep2pt
  \begin{array}{lcllcl}
  p_2 &=& p_1 - p_0 \;, &
  p_3 &=& p_1 - p_0 + q_1 - q_0  \;, \\[1mm]
  p_4 &= &q_1 - q_0  \;, &
  q_3 &=& q_2 - q_1  \;, \\[1mm]
  q_4 &=& q_2 - q_1 - p_0\; .  \hspace{50pt}
  \end{array}
}
For general positions of the arguments, we refer the reader to \cite{2003math......7092N}.


\subsubsection*{Details on the derivation of the four-tachyon correlator}

We now want to apply this formalism to the four-tachyon correlator 
shown in equation \eqref{fourptc}, where in the holomorphic sector the combination 
$\Xi(z_i)={\cal L}({\textstyle {z_{12}\over z_{13}}}) -  
 {\cal L}({\textstyle {z_{12}\over z_{14}}})+ 
 {\cal L}({\textstyle {z_{13}\over z_{14}}}) -
 {\cal L}({\textstyle {z_{23}\over z_{24}}})$ has appeared.
Writing  out the $\mathcal L$ function in terms of the Rogers dilogarithm $L$, we find
\eq{
\label{ampl}
\Xi(z_i)=\hspace{12pt}&L\Bigl(\frac{z_{12}}{z_{13}}\Bigr) - L\Bigl(\frac{z_{12}}{z_{14}}\Bigr) +L\Bigl(\frac{z_{13}}{z_{14}}\Bigr) -L\Bigl(\frac{z_{23}}{z_{24}}\Bigr)  \\[0.1cm]
+&L\Bigl(\frac{z_{13}}{z_{23}}\Bigr) -L\Bigl(\frac{z_{14}}{z_{24}}\Bigr) +L\Bigl(\frac{z_{14}}{z_{34}}\Bigr) -L\Bigl(\frac{z_{24}}{z_{34}}\Bigr)   \\[0.1cm]
+&L\Bigl(\frac{z_{32}}{z_{12}}\Bigr) -L\Bigl(\frac{z_{42}}{z_{12}}\Bigr)
+L\Bigl(\frac{z_{43}}{z_{13}}\Bigr) -L\Bigl(\frac{z_{43}}{z_{23}}\Bigr)\; .
}
Let us then take as above five points in $\mathbb{C}\cup \infty$, 
which are chosen as  $z_1 \dots z_4$ and the point $\infty$.
Recall that our eventual goal is to use the five-term relation in order 
to express  the correlator as 
\begin{equation}
\label{propo}
\Xi(z_i)\simeq L(z) + L\left(\frac{1}{1-z}\right) + L\left(1-\frac{1}{z}\right) + C \;,
\end{equation}
where $z$ is the $SL(2,\mathbb C)$ invariant cross-ratio \eqref{crossratio}
and $C$ is a constant to be determined later. 

We  proceed by  working backwards and writing  down the cross-ratios 
(by omitting every vertex once) corresponding to different permutations of the vertices
$z_1,z_2,z_3,\infty,z_4$, which gives  the desired
arguments in \eqref{propo}. According to the above sketched formalism, 
the other cross-ratios will then have the characteristic form needed for the application of
the five-term relation. 
More concretely, we start with the vertices
$z_1,z_2,z_3,\infty,z_4$ leading to 
\eq{
\label{c1}
  \arraycolsep2pt
  \begin{array}{lcclllclcl}
  [z_1:z_2:z_3:\infty] &=& 
  \displaystyle \frac{z_{32}}{z_{31}} 
  &=:&  x_0 ^A \;, &
  [z_2:z_3:\infty:z_4]  &=& 
  \displaystyle \frac{z_{42}}{z_{43}}  &=:& x_1 ^A  \;,  \\[4mm]
  [z_1:z_3:\infty:z_4] &=&
  \displaystyle  \frac{z_{41}}{z_{43}} &=:& x_2 ^A \;, &
  [z_1:z_2:\infty:z_4] &=& 
  \displaystyle \frac{z_{41}}{z_{42}} &=:& x_3 ^A \;,  \\[4mm]
  [z_1:z_2:z_3:z_4] &=&
  \displaystyle  \frac{z_{32}\, z_{41}}{z_{31}\, z_{42}} &=:&  x_4 ^A\; . \hspace{30pt}
  \end{array}
}
Note that the cross ratio $x_4 ^A$  is  precisely 
the cross ratio in \eqref{crossratio}. Moreover, one finds
$\frac{x_4 ^A}{x_0 ^A} = x_3 ^A$
and the remaining  two characteristic ratios appearing in  the 
five-term relation directly yield $x_1 ^A$ and $x_2 ^A$.
However, only four out of the twelve terms in \eqref{ampl} have  been taken
care off so far. For the other eight, let us first permute the vertices to 
$z_1,z_4,z_2,\infty,z_3$ and  follow the same procedure as before. 
We obtain the following ratios
\eq{
\label{c2}
  \arraycolsep2pt
  \begin{array}{lcclllccll}
  [z_1:z_4:z_2:\infty] &=& 
  \displaystyle \frac{z_{24}}{z_{21}} &=:& x_0 ^B \;, &
  [z_4:z_2:\infty:z_3] &=& 
  \displaystyle \frac{z_{34}}{z_{32}} &=:& x_1 ^B \;,  \\[4mm]
  [z_1:z_2:\infty:z_3] &=& 
  \displaystyle \frac{z_{31}}{z_{32}} &=:& x_2 ^B \;, &
  [z_1:z_4:\infty:z_3] &=& 
  \displaystyle \frac{z_{31}}{z_{34}} &=:& x_3 ^B \;, \\[4mm]
  [z_1:z_4:z_2:z_3] &=& 
  \displaystyle \frac{z_{24}\,z_{31}}{z_{21}\,z_{34}} &=:& x_4 ^B\; . \hspace{30pt}
  \end{array}
}
Note again,  the ratio omitting $\infty$ gives us the argument
$\frac{1}{1-z}$. In the final case we permute the vertices to 
$z_1,z_3,z_4,\infty, z_2$ to find
\eq{
\label{c3}
  \arraycolsep2pt
  \begin{array}{lcclllccll}
  [z_1:z_3:z_4:\infty] &=&
  \displaystyle  \frac{z_{43}}{z_{41}} &=:& x_0 ^C \;, &
  [z_3:z_4:\infty:z_2] &=& 
  \displaystyle \frac{z_{23}}{z_{24}} &=:& x_1 ^C \;,  \\[4mm]
  [z_1:z_4:\infty:z_2] &=&
  \displaystyle  \frac{z_{21}}{z_{24}} &=:& x_2 ^C \;, &
  [z_1:z_3:\infty:z_2] &=&
  \displaystyle  \frac{z_{21}}{z_{23}} &=:& x_3 ^C \;, \\[4mm]
  [z_1:z_3:z_4:z_2] &=&
  \displaystyle  \frac{z_{43}\, z_{21}}{z_{41}\, z_{23}}  &=:& x_4 ^C \;, \hspace{30pt}
  \end{array}
}
and the ratio omitting $\infty$ is the desired argument $1-\frac{1}{z}$.
Comparing the above cross-ratios with the arguments in the four-point 
correlator \eqref{ampl}, we observe  that we can rewrite it in terms of 
the variables $x_k ^L$ defined above
\eq{
\label{ampl2}
\Xi(z_i)=\hspace{12pt}&L\left(1-x_0 ^A\right) - L\Bigl({\textstyle -\frac{x_2^C}{1-x_2^C}}\Bigr) +L\left(1-x_0 ^C\right) -L\left(x_1 ^C\right) \\[0.2cm]
+&L\left(x_2 ^B\right) -L\left(x_3 ^A\right) +L\left(x_2 ^A\right) -L\left(x_1 ^A\right)  \\[0.2cm]
+&L\Bigl({\textstyle \frac{1}{x_3 ^C}}\Bigr) -L\left(x_0^B\right)
+L\Bigl({\textstyle \frac{1}{x_3
  ^B}}\Bigr) -L\left(x_1 ^B\right) \; .
}
To make contact with the extended Rogers dilogarithm and its
five-term relation, we recall  the result of the basic three-point function
\eq{
\label{basic3}
 &\langle {\cal X}^a(z_1,\bar{z}_1)\, {\cal X}^b(z_2,\bar{z}_2)\, {\cal
  X}^c(z_3,\bar{z}_3)\rangle =\\
&\hspace{50pt} \theta^{abc} \left[ 
L\left(\frac{z_{12}}{z_{13}}\right) + L\left(\frac{z_{23}}{z_{21}}\right) +
L\left(\frac{z_{13}}{z_{23}}\right)  \right] + F(z_1,z_2,z_3) - {\rm c.c.} \;,
}
where $F(z_1,z_2,z_3)$ denotes integration ``constants,'' whose
third mixed derivative vanishes.
We then observe that the extended  Rogers dilogarithm
\begin{equation}
R\left(\frac{z_{ij}}{z_{ik}}\right)
 = L\left(\frac{z_{ij}}{z_{ik}}\right) + \frac{i\pi}{2}\left[\, p \log
   \Bigl(\frac{z_{ij}}{z_{ik}}\Bigr) + q \log\Bigl(\frac{z_{jk}}{z_{ik}}\Bigr)
   \right] 
\end{equation}
precisely provides only correction terms which can be interpreted
as such integration constants $F(z_1,z_2,z_3)$.
Thus, taking this choice into account,
we can introduce the  extended Rogers dilogarithm in \eqref{ampl2}.
After some reordering we obtain
\eq{
\label{ampl3}
&\Xi(z_i)=\\
&\arraycolsep2pt
\begin{array}{@{}clclclcl@{}}
&R\left(1-x_0 ^A\,;p_0^A,q_0 ^A\right) &-& 
R\left(x_1^A\,;p_1^A,q_1 ^A\right) &+&
R\left(x_2 ^A\,;p_2^A,q_2^A\right) &-&
R\left(x_3 ^A\,;p_3 ^A,q_3 ^A\right) \\[0.2cm]
-&
R\left(x_0 ^B\,;p_0^B,q_0 ^B\right) &-&
R\left(x_1 ^B\,;p_1 ^B,q_1 ^B\right) &+&
R\left(x_2 ^B\, ;p_2 ^B,q_2 ^B\right) &+&
R\Bigl({\textstyle \frac{1}{x_3 ^B}}\,;p_3 ^B,q_3 ^B\Bigr)   \\[0.2cm]
+&
R\left(1-x_0 ^C\,;p_0 ^C, q_0 ^C\right) &-&
R\left(x_1^C\,;p_1 ^C,q_1 ^C\right) &-&
R\Bigl({\textstyle -\frac{x_2 ^C}{1-x_2 ^C}}\,;p_2 ^C,q_2 ^C\Bigr) &+&
R\Bigl({\textstyle \frac{1}{x_3 ^C}}\,;p_3^C,q_3 ^C\Bigr) .
\end{array}
}
Note that for the extended Rogers dilogarithm the following transformation formulas can be applied
\cite{2003math......6283B} 
\eq{
\label{trafo}
&R(1-x\,;p,q) = -R(x\,;-p,p+q - \epsilon) - \frac{\pi ^2}{6} \;,\\[0.1cm]
&R\Bigl(\frac{1}{x}\,;p,q\Bigr) = -R(x\, ;-p,p+q-\epsilon) -p\frac{\pi ^2}{2}\;, \\[0.1cm]
&R\Bigl(-\frac{x}{1-x}\,;p,q\Bigr) = -R(x\, ;p+q-\epsilon,-q) -\frac{\pi^2}{3}
+ q\frac{\pi^2}{2} \; ,
}
where $\epsilon =\pm1$ for $x$ in  the upper or  lower complex half plane, respectively. It is therefore possible to rewrite the dilogarithm terms purely in terms of the cross ratio parameters determined in \eqref{c1}-\eqref{c3}. If we also take care of the relations between those cross-ratios, up to a constant term we find
\eq{
\label{ampl4}
\Xi(z_i)= 
&-R\left(x_0 ^A\,;-p_0^A,p_0 ^A + q_0 ^A-\epsilon\right) -
 R\Bigl({\textstyle \frac{1-x_0^A}{1-x_4^A}}\,;p_1^A,q_1 ^A\Bigr) \\
&\hspace{40pt}
+R\Bigl({\textstyle \frac{1-(x_0^A)^{-1}}{1-(x_4 ^A)^{-1}}}\,
;p_2^A,q_2^A\Bigr) -R\Bigl({\textstyle \frac{x_4 ^A}{x_0 ^A}}\, ;p_3 ^A,q_3 ^A\Bigr) 
\\[0.2cm]
&-R\left(x_0 ^B\,;p_0^B,q_0 ^B\right) -R\Bigl({\textstyle \frac{1-x_0 ^B}{1-x_4 ^B}}\,;p_1 ^B,q_1 ^B\Bigr) \\
&\hspace{40pt}
+R\Bigl({\textstyle \frac{1-(x_0 ^B)^{-1}}{1-(x_4 ^B)^{-1}}}\, ;p_2 ^B,q_2
^B\Bigr) -R\bigl({\textstyle \frac{x_4 ^B}{x_0 ^B}}\, ;-p_3 ^B,p_3 ^B + q_3 ^B-\epsilon \Bigr) 
\\[0.2cm]
&-R\left(x_0 ^C\, ;-p_0 ^C,p_0 ^C + q_0 ^C-\epsilon\right) 
-R\Bigl({\textstyle \frac{1-x_0^C}{1-x_4 ^C}}\, ;p_1 ^C,q_1 ^C\Bigr) \\
&\hspace{40pt}
+R\Bigl({\textstyle \frac{1-(x_0 ^C)^{-1}}{1-(x_4^C)^{-1}}}\, ;-p_2 ^C,p_2 ^C
+ q_2 ^C-\epsilon\Bigr) 
-R\Bigl({\textstyle \frac{x_4 ^C}{x_0 ^C}}\, ;p_3^C+q_3 ^C -\epsilon,-q_3
^C\Bigr).
}
Imposing now the flattening conditions separately on $p_i ^{A,B,C}, q_j ^{A,B,C}$ in such a way that 
\begin{equation}
p_4 ^{A,B,C} = q_4 ^{A,B,C} = 0 \;,
\end{equation}
(this is possible since in every region the flattening condition determines only five of the $10$ parameters involved in the five-term relation in terms of the remaining ones), we finally get 
\eq{
\label{final}
\Xi(z_i)=-R(x_4^A;0,0) - R(x_4 ^B;0,0) - R(x_4 ^C;0,0) + C' = -{\cal L}(z) + C' \;,
}
which is equal to \eqref{propo} up to a constant.
We can now uniquely fix this constant by requiring crossing
symmetry of the holomorphic part of the four-point function.
Due to the fundamental relation \eqref{complfund}, one obtains
$C' = {3\over 2}\, L(1)=\frac{\pi^2 }{4}$. 
To summarize, employing the five-term relation for the extended Rogers
dilogarithm, it is possible to chose the integration constants in our 
basic correlator such that 
\eq{
\label{finalb}
{\cal L}({\textstyle {z_{12}\over z_{13}}}) -  
 {\cal L}({\textstyle {z_{12}\over z_{14}}})+ 
 {\cal L}({\textstyle {z_{13}\over z_{14}}}) -
 {\cal L}({\textstyle {z_{23}\over z_{24}}})=
 -{3\over 2}L(1)+ {\cal L}(X)\; ,
}
so that the four point tachyon correlator is both crossing symmetric 
and depends only on the cross ratio 
$X=1-z$ (and is therefore $SL(2,\mathbb{C})$-invariant).


\subsubsection*{Relation between ${\cal L}(z)$ and $L(z)$}

Let us finally elaborate on the relation between the $\mathcal L$ function and the usual
Rogers dilogarithm $L$.
For the Taylor expansion of ${\cal L}(z)$ around $z=0$ it is useful
to first relate the sum in ${\cal L}(z)$ to $L(z)$.
We will show that the following relation holds
\begin{equation}
  \label{3L}
  L(z) + L\left(1-\frac{1}{z}\right) + L\left(\frac{1}{1-z}\right) 
  = 3\,L(z) \pm \frac{i\pi}{2} \log\bigl(z(1-z)\bigr) \;.
\end{equation} 
Note that here $z \in \mathbb{C} \setminus \mathbb{R}$ and the plus sign holds for $\textrm{Im}(z) < 0$ while the minus sign holds for $\textrm{Im}(z) >0$.
For the proof of \eqref{3L} we recall from equation \eqref{euler} that the arguments of the Euler dilogarithm transform as 
\eq{
\textrm{Li}_2\Bigl( \frac{z-1}{z}\Bigr) &= \textrm{Li}_2(z) -\frac{\pi^2}{6} - \frac{1}{2} \log ^2 (z) + \log(1-z)\log(z) \;, \\
\textrm{Li}_2\Bigl(\frac{1}{1-z}\Bigr) &= \textrm{Li}_2(z) + \frac{\pi^2}{6} + \log(-z)\log(1-z) -\frac{1}{2} \log^2(1-z) \;.
} 
We therefore find
\eq{
  \label{3Lfast}
  &L(z) + L\left(1-\frac{1}{z}\right) + L\left(\frac{1}{1-z}\right)\\
  & \hspace{80pt}= 3\,L(z) +\frac{1}{2} \log(-z)\log(1-z) - \frac{1}{2}\log(z-1)\log(z) \;.
}
To rewrite the logarithmic terms with negative argument, one has to distinguish  two cases depending on the sign of the imaginary part of $z$
\eq{
\log(-z) &= \log(z) - i\pi  \qquad{\rm for}\qquad \textrm{Im}(z) > 0 \;, \\
\log(-z) &= \log(z) + i \pi \qquad{\rm for}\qquad \textrm{Im}(z) < 0 \;.
}
Using this distinction in equation \eqref{3Lfast}, one arrives at the desired result \eqref{3L}.


\subsection{Details on correlation functions}
\label{app_twopt}

In this appendix, we present some details on the computation of correlation functions which have appeared in the main text.


\subsubsection*{Correction to two-point function}

We consider the correction to the two-point function of two fields $X^a(z,\ov z)$ at second order in the $H$-flux, and for convenience we recall formula \eqref{tpf_corr_78}
\eq{
  \hspace{185pt}&\hspace{-185pt}
  \delta_2 \bigl\langle  X^a(z_1,\ov z_1) X^b(z_2,\ov z_2)  \bigr\rangle
   = \frac{1}{2\,(6\pi \alpha')^2} \: H_{mno} \, H_{pqr} 
   \int_{\Sigma} d^2 w_1  \int_{\Sigma} d^2 w_2  \\[0.1cm]
 \bigl\langle :\! X^a(z_1,\ov z_1) X^b(z_2,\ov z_2)\!:   
 &:\! X^m(w_1,\ov w_1)\, \partial X^n(w_1) \,\ov\partial X^o(\ov w_1) \!:  \\
 &:\! X^p(w_2,\ov w_2)\hspace{4.5pt} \partial X^q(w_2) \hspace{3.5pt}
 \ov\partial X^r(\ov w_2) \!: \bigr\rangle_{0} \,.
}
The correlator is evaluated using Wick contractions and can be simplified using partial integration. For the latter one should note  that the integrals have to be regularized by cutting out of the integration region a small disc of radius $\epsilon\ll1$ specified by $|\omega_1-\omega_2|<\epsilon$. One is then left with computing
\eq{
  &\delta_2 \bigl\langle  X^a(z_1,\ov z_1) X^b(z_2,\ov z_2)  \bigr\rangle \\
  &\hspace{45pt}
  = -\frac{ {\alpha'}^2}{64 \pi^2} \: H^a{}_{pq}\, H^{bpq}
   \int d^2 w_1 \, d^2 w_2 \:
   \frac{1}{z_1-w_1}\:\frac{1}{\ov z_2-\ov w_2}\:\frac{1}{|w_1-w_2|^2}\;.
}
Let us now rewrite this expression in order to apply a variant of the inhomogeneous Cauchy formula. In particular, we have
\eq{
  \label{app_some_res_837}
  &\delta_2 \bigl\langle  X^a(z_1,\ov z_1) X^b(z_2,\ov z_2)  \bigr\rangle \\
  & \hspace{20pt}
  = -\frac{ {\alpha'}^2}{64 \pi^2} \: H^a{}_{pq}\, H^{bpq}
  \int d^2 w_2 \, \frac{1}{\ov z_2-\ov w_2}
  \int d^2 w_1\,   \frac{1}{w_1-w_2} \:\ov\partial_{\ov w_1} \left(
  \frac{\log |w_1-w_2|^2}{z_1-w_1}\right) .
}
For the second integral we apply the regularization procedure mentioned above and remove the disc $D_\epsilon(w_2)$ of radius $\epsilon\ll1$ from the $w_1$-plane. More concretely, we compute
\begin{align}
  \nonumber
  \int d^2 w_1\,   \frac{1}{w_1-w_2} \:\ov\partial_{\ov w_1} \left(
  \frac{\log |w_1-w_2|^2}{z_1-w_1}\right) 
   &=
   \oint_{\partial D_\epsilon(w_2)}\!\! d w_1 \,
   \frac{1}{w_1-w_2} \:  \frac{\log |w_1-w_2|^2}{z_1-w_1} \\[2mm]
   &= 4\pi i\, \frac{\log\epsilon}{z_1-w_2} \; .
\end{align}
Using this result in \eqref{app_some_res_837}, we obtain the expression
\eq{
  \delta_2 \bigl\langle  X^a(z_1,\ov z_1) X^b(z_2,\ov z_2)  \bigr\rangle 
  = -\frac{ i\, {\alpha'}^2 \log \epsilon}{16 \pi} \: H^a{}_{pq}\, H^{bpq}
  \int d^2 w_2 \,   \frac{1}{z_1-w_2}\:\frac{1}{\ov z_2-\ov w_2} \;,
}
for which we apply the above procedure once more. This leads to the final expression
\eq{
  \label{app_some_res_838}
  \delta_2 \bigl\langle  X^a(z_1,\ov z_1) X^b(z_2,\ov z_2)  \bigr\rangle 
  = \frac{{\alpha'}^2}{8} \: H^a{}_{pq}\, H^{bpq}
   \log|z_1-z_2|^2 \,\log \epsilon\;.
}


\subsubsection*{$N$-tachyon correlator}
\label{app_Ntachy}

We now turn to the evaluation of a correlator involving $N$ tachyon vertex operators of the form
\eq{
  {\cal V}^-_i \equiv {\cal V}_{p_i}(z_i,\ov z_i)= :\! \exp \bigr(\, i\hspace{0.5pt} p_i 
  \cdot {\cal X}(z_i,\ov z_i) \bigl) : \;,
}
where $i=1,\ldots, N$ labels the different operators and where 
we again employ the short-hand notation $p\cdot {\cal X}=  p_a {\cal X}^a$.
Using formula \eqref{master_01}, up to first order in the flux the $N$-tachyon amplitude in the $H$-flux background can be expanded as
\eq{
  \label{some_res_42}
 \bigl\langle {\cal V}_1 \ldots {\cal V}_N \bigr\rangle^- = 
 \bigl\langle {\cal V}_1 \ldots {\cal V}_N \bigr\rangle^-_0   
 - \bigl\langle {\cal V}_1 \ldots {\cal V}_N \,\mathcal S_1 \bigr\rangle^-_0
 + \mathcal O (H^2) \;.
}
Let us consider the second term involving $\mathcal S_1$ in more detail. Since the perturbation $\mathcal S_1$ is already linear in the flux $H$, we can replace the perturbed vertex operators $\mathcal V_i^-$ by the ones of the free theory $V_i^-$, defined in terms of the free field $X^a(z,\ov z)$.
In this case, we can employ standard techniques to evaluate the correlator and find (at linear order in the flux)
\eq{
  \label{some_res_6}
  &\bigl\langle {\cal V}_1 \ldots {\cal V}_N \:(-\mathcal S_1) \bigr\rangle^-_0 
  = \bigl\langle {V}_1 \ldots {V}_N \bigr\rangle^-_0  \times \\
  &\hspace{60pt}  \sum_{1\leq i <  j <  k\leq N}
 (-i)\:p_{i, a} \,p_{j,b}\, p_{k,c} \, \bigl\langle X^a(z_i,\ov z_i) X^b(z_j,\ov z_j) X^c(z_k,\ov z_k) 
  \bigr\rangle^- \;.
}
Concerning the first term in \eqref{some_res_42}, we recall the definition of the corrected field as
${\cal X}^a(z,\ov z) = X^a(z,\ov z) + {\textstyle \frac{1}{2}}\hspace{0.5pt}H^a{}_{bc} \,  X^b_L(z) \, X^c_R(\ov z)$, and expand the exponentials in the vertex operators up to linear order in $H$.
Combining this contribution with \eqref{some_res_6} leads to the following result (up to linear order in the flux)
\eq{
  \label{app_Ntachpre}
  \bigl\langle {\cal V}_1 \ldots {\cal V}_N \bigr\rangle^- &= 
  \bigl\langle {V}_1 \ldots {V}_N \bigr\rangle^-_0 \:\biggl[ 1 +
  \!\!\!\! \sum_{1\leq i <  j <  k\leq N}  \!\!\!\!
 (-i)\:p_{i, a} \,p_{j,b}\, p_{k,c} \times \\[-0.2cm]
 & \hspace{143pt}
 \bigl\langle {\cal X}^a(z_i,\ov z_i)\, {\cal X}^b(z_j,\ov z_j) \,{\cal X}^c(z_k,\ov z_k) 
  \bigr\rangle^- \biggr] \\[0.1cm]
 &=
 \bigl\langle {V}_1 \ldots {V}_N \bigr\rangle^-_0 \:\biggl[ 1 -i\hspace{0.5pt} 
   \theta^{abc}\!\!\!\! \sum_{1\leq i<j<k\leq N}  \!\!\!\! 
    p_{i ,a} \,p_{j,b}\, p_{k,c}  
   \Bigl[ {\cal L}\bigl({\textstyle \frac{z_{ij}}{ z_{ik}}}\bigr)
  -  {\cal L}\bigl({\textstyle \frac{\ov z_{ij}}{ \ov z_{ik}}}\bigr)
  \Bigr] \biggr] \;,
}
where we inserted the expression for the basic three-point function 
\eqref{three-point_01}. 
The computation for  vertex operators describing winding states
${\cal V}^+_i$ is analogous and only leads to the usual sign flip
between the holomorphic and anti-holomorphic contributions.
With  $[\ldots ]_\theta$ indicating that the exponential
is to be expanded only up to linear order in $\theta$,
we then arrive at
\eq{
  \label{app_Ntach}
  \bigl\langle {\cal V}_1 \ldots {\cal V}^\mp_N \bigr\rangle
  &= 
  \bigl\langle {V}_1 \ldots {V}_N \bigr\rangle^\mp_0 \: \exp \biggl[  \:
   -i \hspace{0.5pt} \theta^{abc} \!\!\!\! \sum_{1\leq i<j<k\leq N} \!\!\!\!  p_{i ,a} \,p_{j,b}\, p_{k,c}  
   \Bigl[ {\cal L}\bigl({\textstyle \frac{z_{ij}}{ z_{ik}}}\bigr)
  \mp  {\cal L}\bigl({\textstyle \frac{\ov z_{ij}}{ \ov z_{ik}}}\bigr)
  \Bigr] \biggr]_{\theta} \;.
}


\clearpage
\nocite{*}
\bibliography{rev}  
\bibliographystyle{utphys}


\end{document}